\documentclass{lmcs}
\pdfoutput=1

\usepackage{lastpage}
\lmcsdoi{21}{2}{14}
\lmcsheading{}{\pageref{LastPage}}{}{}%
{Nov.~02,~2021}{May~16,~2025}{}

\usepackage[utf8]{inputenc} 
\usepackage[framemethod=TikZ]{mdframed}
\usepackage{calc}
\usepackage{tcolorbox}
\usepackage{tabto}
\usepackage{amssymb}
\usepackage{graphicx}   
\usepackage{xspace}
\usepackage{verbatim}   
\usepackage{xcolor}
\usepackage{amsmath}
\usepackage{makecell}
\usepackage{tikz,pgfplots}
\usepackage{relsize}
\usepackage{booktabs}
\usepackage{tikz-qtree}
\usepackage{subcaption}
\usepackage{multirow}
\usetikzlibrary{arrows.meta}
\usetikzlibrary{fit,shapes,trees,shapes.geometric}
\usetikzlibrary{patterns,decorations.pathreplacing,calc}
\usetikzlibrary{matrix, positioning, arrows}
\usetikzlibrary{chains,shapes.multipart}
\usetikzlibrary{shapes,calc}
\usetikzlibrary{automata}
\usepackage[linesnumbered,algoruled, lined, noend]{algorithm2e}
\usepackage{footnote}
\def\mystrut{\vrule height 1em depth 0.4em width 0em} 
\def\mystrutcopy{\vrule height 0.6em depth 0.25em width 0em } 
  
\newcommand{\highlight}[1]{{\color{black} {#1}}}

\DeclareRobustCommand\oldtt[1]{{\mbox{\bfseries \texttt{#1}}}}

\newsavebox{\bullmagenta}
\newsavebox{\bullolive}
\newsavebox{\bullblue}
\newsavebox{\bullteal}
\newsavebox{\bullbrown}
\newsavebox{\bullorange}
\sbox\bullmagenta{\tikz{\draw[magenta,fill=magenta] circle (.3ex);}}
\sbox\bullolive{\tikz{\draw[olive,fill=olive] circle (.3ex);}}
\sbox\bullblue{\tikz{\draw[blue!30,fill=blue!30] circle (.3ex);}}
\sbox\bullteal{\tikz{\draw[teal,fill=teal] circle (.3ex);}}
\sbox\bullbrown{\tikz{\draw[brown,fill=brown] circle (.3ex);}}
\sbox\bullorange{\tikz{\draw[orange,fill=orange] circle (.3ex);}}

\DeclareRobustCommand\mytikzcell{\tikz[baseline=-2.5pt]	\tikzset{b2block/.style = {rectangle split, rectangle split parts=1,
			draw, rectangle split horizontal=false,inner ysep=0pt, line width=0.05em, text width=6em, align=center, rectangle split part fill={blue!30}}} \node[b2block] {
		\nodepart{one} {\mystrutcopy \tiny $\mathtt{\langle 1, [30, 40], \bot \rangle}$ \enskip {$3$}}};}

\DeclareRobustCommand\mytikzcellcopy{\tikz[baseline=-3pt] 			\tikzset{b2block/.style = {rectangle split, rectangle split parts=1,
			draw, rectangle split horizontal=false,inner ysep=0pt, line width=0.05em, text width=6em, align=center, rectangle split part fill={blue!30}}} \node[b2block] {
		\nodepart{one} {\mystrutcopy \tiny $\mathtt{\langle 1, [30, 40], 11 \rangle}$ \enskip {$3$}}};}	

\DeclareRobustCommand\myrootcell{\tikz[baseline=-2pt] 			\tikzset{b2block/.style = {rectangle split, rectangle split parts=1,
			draw, rectangle split horizontal=false,inner ysep=0pt, line width=0.05em, text width=6em, align=center, rectangle split part fill={white}}} \node[b2block] {
		\nodepart{one} {\mystrutcopy \tiny $\mathtt{\langle 1, [10], \bot \rangle}$ \quad {$4$}}};}	

\DeclareRobustCommand\mybthreecell{\tikz[baseline=-3pt] 			\tikzset{b2block/.style = {rectangle split, rectangle split parts=1,
			draw, rectangle split horizontal=false,inner ysep=0pt, line width=0.05em, text width=4.5em, align=center, rectangle split part fill={orange!60}}} \node[b2block] {
		\nodepart{one} {\mystrutcopy \tiny $\mathtt{\langle 1, [], \bot \rangle}$ \quad {$1$}}};}		
	
\DeclareRobustCommand\mybthreecellnext{\tikz[baseline=-3pt] 			\tikzset{b2block/.style = {rectangle split, rectangle split parts=1,
			draw, rectangle split horizontal=false,inner ysep=0pt, line width=0.05em, text width=4.5em, align=center, rectangle split part fill={orange!60}}} \node[b2block] {
		\nodepart{one} {\mystrutcopy \tiny $\mathtt{\langle 2, [], \bot \rangle}$ \quad {$4$}}};}			

\DeclareRobustCommand\mybtwocell{\tikz[baseline=-3pt] 			\tikzset{b2block/.style = {rectangle split, rectangle split parts=1,
			draw, rectangle split horizontal=false,inner ysep=0pt, line width=0.05em, text width=6em, align=center, rectangle split part fill={blue!30}}} \node[b2block]{ \nodepart{one} {\mystrutcopy \tiny $\mathtt{\langle 1, [31, 40], \bot \rangle}$ \enskip {$6$}}};}	
		
\DeclareRobustCommand\mybtwocellcopy{\tikz[baseline=-3pt] 			\tikzset{b2block/.style = {rectangle split, rectangle split parts=1,
					draw, rectangle split horizontal=false,inner ysep=0pt, line width=0.05em, text width=6em, align=center, rectangle split part fill={blue!30}}} \node[b2block]{ \nodepart{one} {\mystrutcopy \tiny $\mathtt{\langle 1, [30, 41], \bot \rangle}$ \enskip {$7$}}};}	
				
\DeclareRobustCommand\mybtwocellcopyfinal{\tikz[baseline=-3pt] 			\tikzset{b2block/.style = {rectangle split, rectangle split parts=1,
			draw, rectangle split horizontal=false,inner ysep=0pt, line width=0.05em, text width=6em, align=center, rectangle split part fill={blue!30}}} \node[b2block]{ \nodepart{one} {\mystrutcopy \tiny $\mathtt{\langle 1, [31, 41], \bot \rangle}$ \enskip {$10$}}};}					

\DeclareRobustCommand\mybonecell{\tikz[baseline=-3pt] 			\tikzset{b2block/.style = {rectangle split, rectangle split parts=1,
			draw, rectangle split horizontal=false,inner ysep=0pt, line width=0.05em, text width=6em, align=center, rectangle split part fill={white}}} \node[b2block]{ \nodepart{one} {\mystrutcopy \tiny $\mathtt{\langle 2, [10], \bot \rangle}$ \enskip {$5$}}};}		

\DeclareRobustCommand\mybonecellcopy{\tikz[baseline=-3pt] 			\tikzset{b2block/.style = {rectangle split, rectangle split parts=1,
			draw, rectangle split horizontal=false,inner ysep=0pt, line width=0.05em, text width=4.5em, align=center, rectangle split part fill={white}}} \node[b2block]{ \nodepart{one} {\mystrutcopy \tiny $\mathtt{\langle 2, [11], \bot \rangle}$ \enskip {$5$}}};}

\newcommand{\eat}[1]{}
\newcommand{\mathsout}[1]
{\bgroup\mathchoice
	{\sbox0{$\displaystyle{#1}$}%
		\usebox0\hspace{-\wd0}%
		\rule[0.5\ht0-0.5\dp0-.5pt]{\wd0}{1pt}}%
	{\sbox0{$\textstyle{#1}$}%
		\usebox0\hspace{-\wd0}%
		\rule[0.5\ht0-0.5\dp0-.5pt]{\wd0}{1pt}}%
	{\sbox0{$\scriptstyle{#1}$}%
		\usebox0\hspace{-\wd0}%
		\rule[0.5\ht0-0.5\dp0-.5pt]{\wd0}{1pt}}%
	{\sbox0{$\scriptscriptstyle{#1}$}%
		\usebox0\hspace{-\wd0}%
		\rule[0.5\ht0-0.5\dp0-.5pt]{\wd0}{1pt}}%
	\egroup}

\newcommand{\cut}[1]{}   
\makeatletter
\newcommand{\mytag}[2]{%
	\text{#1}%
	\@bsphack
	\protected@write\@auxout{}%
	{\string\newlabel{#2}{{#1}{\thepage}}}%
	\@esphack
}
\makeatother

\newenvironment{packed_enum}{
	\begin{enumerate}
		\setlength{\itemsep}{1pt}
		\setlength{\parskip}{0pt}
		\setlength{\parsep}{0pt}
	}
	{\end{enumerate}}

\newcommand{\ie}{{\em i.e.}\xspace}

\newcommand{\out}{\textsf{output}}

\newlength\myboxwidth

\definecolor{code}{RGB}{230,230,230}

\newcommand{\introparagraph}[1]{\subparagraph{{\bf #1}}}  

\usepackage{aliascnt} 

\newtheorem{claim}[thm]{Claim}

\providecommand{\bx}[0]{\mathbf{x}}

\providecommand{\bu}[0]{\mathbf{u}}

\providecommand{\mk}[0]{k}

\providecommand{\mH}[0]{\mathcal{H}}

\providecommand{\mB}[0]{\mathcal{B}}

\providecommand{\mbroot}[0]{\mB_{\texttt{r}}}

\providecommand{\mL}[0]{\mathcal{L}}

\providecommand{\tOUT}[0]{\texttt{OUT}}
\providecommand{\dia}[0]{\textsf{dia}}

\providecommand{\prnt}[1]{\texttt{p}({#1})}
\providecommand{\key}[1]{\texttt{key}({#1})}
\providecommand{\val}[1]{\texttt{value}({#1})}
\providecommand{\subtree}[1]{\mB^\prec_{#1}}

\providecommand{\qQ}[0]{\mathfrak{Q}}

\providecommand{\rank}[1]{\texttt{rank}({#1})}
\providecommand{\mrf}[0]{\textnormal{\texttt{rank}}}
\providecommand{\rf}[0]{\mrf}
\providecommand{\mfhw}[0]{\textnormal{\texttt{fhw}}}
\providecommand{\msubw}[0]{\textnormal{\texttt{subw}}}

\providecommand{\fhw}[0]{\mfhw}
\providecommand{\subw}[0]{\msubw}

\providecommand{\panda}[0]{\texttt{PANDA}}
\providecommand{\fhwob}[0]{\textnormal{\texttt{fhw-lex}}}
\providecommand{\edges}[0]{\mathcal{E}}
\providecommand{\nodes}[0]{\mathcal{V}}
\providecommand{\vars}[1]{\textsf{vars}(#1)}

\providecommand{\domain}[0]{\mathbf{dom}}

\providecommand{\htree}[0]{\mathcal{T}}
\providecommand{\bag}[0]{\mathcal{B}}
\providecommand{\fhw}[1]{\mathsf{fhw}(#1)}

\providecommand{\eat}[1]{}

\providecommand{\mnext}[0]{\mathsf{next}}

\newcommand{\hlrone}[1]{{\color{black} {#1}}}
\newcommand{\hlrtwo}[1]{{\color{black} {#1}}}

\tikzset{
	ncbar angle/.initial=90,
	ncbar/.style={
		to path=(\tikztostart)
		-- ($(\tikztostart)!#1!\pgfkeysvalueof{/tikz/ncbar angle}:(\tikztotarget)$)
		-- ($(\tikztotarget)!($(\tikztostart)!#1!\pgfkeysvalueof{/tikz/ncbar angle}:(\tikztotarget)$)!\pgfkeysvalueof{/tikz/ncbar angle}:(\tikztostart)$)
		-- (\tikztotarget)
	},
	ncbar/.default=0.25cm,
}

\tikzset{square left brace/.style={ncbar=0.25cm}}
\tikzset{square right brace/.style={ncbar=-0.25cm}}

\tikzset{round left paren/.style={ncbar=0.25cm,out=120,in=-120}}
\tikzset{round right paren/.style={ncbar=0.25cm,out=60,in=-60}}

\begin{document}
	
	\title{Ranked Enumeration of Conjunctive Query Results}

    \thanks{We are grateful to the reviewers for a careful reading of the manuscript and their feedback.}
    
	\author{Shaleen Deep\lmcsorcid{0000-0003-2342-4060}}
	\address{Department of Computer Sciences, University of Wisconsin-Madison, Madison, Wisconsin, USA}
	\email{shaleen@cs.wisc.edu, paris@cs.wisc.edu}

	
	\author{Paraschos Koutris\lmcsorcid{0000-0001-6309-1702}}

	
	\maketitle
	
	\begin{abstract}
We study the problem of enumerating answers of Conjunctive Queries ranked according to a given ranking function. Our main contribution is a novel algorithm with small preprocessing time, logarithmic delay, and non-trivial space usage during execution. To allow for efficient enumeration, we exploit certain properties of ranking functions that frequently occur in practice. To this end, we introduce the notions of {\em decomposable} and {\em compatible} (w.r.t. a query decomposition) ranking functions, which allow for partial aggregation of tuple scores in order to efficiently enumerate the output. We complement the algorithmic results with lower bounds that justify why restrictions on the structure of ranking functions are necessary. Our results extend and improve upon a long line of work that has studied ranked enumeration from both a theoretical and practical perspective.
\end{abstract}

	\section{Introduction}
\label{sec:intro}

For many data processing applications, enumerating query results according to an order given by a ranking function is a fundamental task. For example, \cite{yang2018any, chang2015optimal} consider a setting where users want to extract the top patterns from an edge-weighted graph, where the rank of each pattern is the sum of the weights of the edges in the pattern. Ranked enumeration also occurs in \oldtt{SQL} queries with an \oldtt{ORDER BY} clause~\cite{qi2007sum, ilyas2004rank}.
In the above scenarios, the user often wants to see the first $k$ results in the query as quickly as possible, but the value of $k$ may not be predetermined. Hence, it is critical to construct algorithms that can output the first tuple of the result as fast as possible, and then output the next tuple in the order with a very small {\em delay}.
In this article, we study the algorithmic problem of enumerating the result of a Conjunctive Query (CQ, for short) against a relational database where the tuples must be output in order given by a ranking function.

The simplest way to enumerate the output is to materialize the result $\tOUT$ and sort the tuples based on the score of each tuple. Although this approach is conceptually simple, it requires that $\vert \tOUT \vert$ tuples are materialized; moreover, the time from when the user submits the query to when she receives the first output tuples is $\Omega (\vert \tOUT \vert \cdot \log \vert \tOUT \vert)$. Further, the space and delay guarantees do not depend on the number of tuples that the user wants to actually see. 
 More sophisticated approaches to this problem construct optimizers that exploit properties such as the monotonicity of the ranking function, allowing for join evaluation on a subset of the input relations (see~\cite{ilyas2008survey} and references within). In spite of the significant progress, all of the known techniques suffer from large worst-case space requirements, no dependence on $k$, and provide no non-trivial guarantees on the delay during enumeration, {with the exception of a few cases where the ranking function is of a special form}. Fagin et al.~\cite{fagin2003optimal} initiated a long line of study related to aggregation over {\em sorted lists}. However,~\cite{fagin2003optimal} and subsequent works also suffer from the above mentioned limitations as we do not have the materialized output $Q(D)$ that can be used as sorted lists.

In this article, we construct algorithms that remedy some of these issues. Our algorithms are divided into two phases: the {\em preprocessing phase}, where the system constructs a data structure that can be used later and the {\em enumeration phase}, when the results are generated. All of our algorithms aim to minimize the time of the preprocessing 
phase, and guarantee a {\em logarithmic delay} ${O}(\log |D|)$ during enumeration. Although we cannot hope to
perform efficient ranked enumeration for an arbitrary ranking function, we show that our techniques apply for most ranking 
functions of practical interest, including lexicographic ordering, and sum (also product or max) of weights of input tuples among others.

\begin{exa} \label{ex:intro}
Consider a weighted graph $G$, where an edge $(a,b)$ with weight $w$ is represented by the relation $R(a,b,w)$.
Suppose that the user is interested in finding the (directed) paths of length 3 in the graph with the lowest score, where
the score is a (weighted) sum of the weights of the edges. The user query in this case can be specified as:
$ Q(x,y,z,u, w_1, w_2, w_3, ) = R(x,y,w_1), R(y,z, w_2), R(z,u,w_3)$
where the ranking of the output tuples is specified for example by the score $5w_1 + 2w_2 + 4w_3$. 
If the graph has $N$ edges, the na\"ive algorithm that computes and ranks all tuples needs $\Omega(N^2 \log N)$ preprocessing time. We show that it is possible to design an algorithm with $O(N)$ preprocessing time, such that
the delay during enumeration is ${O}(\log N)$. This algorithm outputs the first $k$ tuples by
materializing $O(N+k)$ data, even if the full output is much larger. 
\end{exa}

The problem of ranked enumeration for CQs has been studied both theoretically~\cite{kimelfeld2006incrementally, cohen2007incremental, DBLP:journals/tods/OlteanuZ15,tziavelis13optimal} and practically~\cite{yang2018any, chang2015optimal, bakibayev2012fdb}. 
Theoretically,~\cite{kimelfeld2006incrementally} establishes the tractability of enumerating answers in sorted order with polynomial delay (combined complexity), albeit with suboptimal space and delay factors for two classes of ranking functions. \cite{yang2018any} presents an anytime enumeration algorithm restricted to acyclic queries on graphs that uses $\Theta(|\tOUT| + |D|)$ space in the worst case, has a $\Theta(|D|)$ delay guarantee, and supports only simple ranking functions.
As we will see, both of these guarantees are suboptimal and can be improved upon. 

Ranked enumeration has also been studied for the class of lexicographic orderings. In~\cite{bagan2007acyclic}, the authors show that {\em free-connex acyclic CQs} can be enumerated in constant delay after only linear time preprocessing. Here, the lexicographic order is chosen by the algorithm and not the user. 
Factorized databases~\cite{bakibayev2012fdb, DBLP:journals/tods/OlteanuZ15} can also support constant delay ranked enumeration, but only when the lexicographic ordering agrees with the order of the query decomposition. In contrast, our results imply that we can achieve a logarithmic delay with the same preprocessing time for {\em any} lexicographic order. \hlrtwo{Concurrent work~\cite{tziavelis13optimal} has also considered ranked enumeration from a theoretical and practical perspective. Our work recovers some of the theoretical results presented in~\cite{tziavelis13optimal}. Further, we consider a broader set of ranking functions\eat{\footnote{The reader can find a more detailed comparison in Section~\ref{sec:related}.}} compared to all prior works along with new lower bounds.}

\smallskip
\introparagraph{Our {Contributions}} In this work, we show how to obtain logarithmic delay guarantees with small preprocessing time  for ranking results of full (projection-free) CQs. We summarize our technical contributions below:

\begin{enumerate}

\item Our main contribution (Theorem \ref{thm:main}) is a novel algorithm that uses query decomposition techniques in conjunction with structure of the ranking function. The preprocessing phase sets up priority queues that maintain partial tuples at each node of the decomposition. During the enumeration phase, the algorithm materializes the output of the subquery formed by the subtree rooted at each node of the decomposition {\em on-the-fly}, in sorted order according to the ranking function. In order to define the rank of the partial tuples, we require that the ranking function can be \emph{decomposed} with respect to the particular decomposition at hand. Theorem~\ref{thm:main} then shows that with $O(|D|^\fhw)$ preprocessing time, where \fhw\ is the {\em fractional hypertree width} of the decomposition, we can enumerate with delay ${O}(\log |D|)$. We then discuss how to apply our main result to  commonly used classes of ranking functions. Our work thoroughly resolves an open problem stated at the Dagstuhl Seminar 19211~\cite{boros_et_al} on ranked enumeration (see Question $4.6$).

\item We propose two extensions of Theorem~\ref{thm:main} that improve the preprocessing time to $O(|D|^\subw)$, a polynomial improvement over Theorem~\ref{thm:main} where \subw\ is the {\em submodular width} of the query $Q$. The result is based on a simple but powerful application of the main result that can be applied to any full UCQ $Q$ combined with the \panda\ algorithm proposed by Abo Khamis et al.~\cite{abo2017shannon}.

\item Finally, we show lower bounds (conditional and unconditional) for our algorithmic results.
In particular, we show that subject to a popular conjecture, the logarithmic factor in delay cannot be removed. Additionally, we show that for two particular classes of ranking functions, we provide simple properties over the hypergraph that characterize whether it is possible to achieve logarithmic delay with linear preprocessing time for a large class of fully acyclic CQs. 
\end{enumerate}
\noindent 
This article is the full version of a conference publication~\cite{deep2021ranked}. We have added all of the proofs and intermediate results that were excluded from the paper. In particular, we have added the full proof of our main result -- ranked enumeration of full CQs (\autoref{thm:main}). Additionally, we have also added the full algorithm for the the extensions in~\autoref{sec:extension}. We have reworked the example for the main result and added a more detailed discussion to improve the exposition. \hlrone{In~\autoref{sec:lowerbound}, we present dichotomy results for \emph{graph} queries (i.e., queries over binary relations)\footnote{We also amend an error in the conference publication version.}.} Finally, we have added a brief discussion in the conclusion regarding the extension of our results to the dynamic setting based on discussion with other community members at ICDT 2021. The remainder of the article is organized as follows. In the next section, we {present} the preliminaries and basic notation. Section~\ref{sec:proof} shows the first main result (\autoref{thm:main}), which is subsequently used as a building block in~\autoref{sec:extension} for the second main result (\autoref{thm:ucq} and~\autoref{thm:main:subw}). Lower bounds are presented in~\autoref{sec:lowerbound} and the related work in~\autoref{sec:related}. We conclude with a list of open problems in~\autoref{sec:conclusion}.

	\section{Problem Setting}
\label{sec:framework}

In this section we present the basic notions and terminology, and then discuss our framework.

\subsection{Conjunctive Queries}
\label{subsec:cq}

We will focus on the class of {\em Conjunctive Queries (CQs)}, which are expressed as
$ \label{eq:q}
Q(\mathbf{y}) = R_1(\bx_1), R_2(\bx_2), \ldots, R_n(\bx_n) 
$
Here, the symbols $\mathbf{y},\bx_1, \dots, \bx_n$ are vectors that contain {\em variables} 
or {\em constants}, the atom $Q(\mathbf{y})$ is the {\em head} of the query, and the atoms
$R_1(\bx_1), R_2(\bx_2), \ldots, R_n(\bx_n)$ form the {\em body}. 
The variables in the head are a subset of the variables that appear in the body. \highlight{We use $\vars{Q}$ to denote the set of all variables in $Q$, i.e., $\bx_1 \cup \dots \cup \bx_n$.} A CQ  is 
{\em full} if every variable in the body appears also in the head, and it is {\em boolean} 
if the head contains no variables, \ie it is of the form $Q()$. \highlight{If $\bx_i \subseteq \bx_j$, we can join $R_i(\bx_i)$ and $R_j(\bx_j)$, followed by removing atom $R_i$ from the query.}

We will typically use the symbols 
$x,y,z,\dots$ to denote variables, and $a,b,c,\dots$ to denote constants.
We use $Q(D)$ to denote the result of the full CQ $Q$ over input database $D$.
A {\em valuation} $\theta$ over a set $V$ of variables is a total function that maps
each variable $x \in V$ to a value $\theta(x) \in \domain$, where $\domain$ is a domain
of constants. We will often use $\domain(x)$ to denote the constants that the valuations
over variable $x$ can take. 
It is implicitly understood that a valuation is the identity function on constants.
If $U \subseteq V$, then $\theta[U]$ denotes the restriction of $\theta$ to $U$. \highlight{An \textit{answer} \hlrone{to a full CQ $Q$} is a tuple $\theta(\vars{Q})$ which is a mapping from $\vars{Q}$ to $\domain$ such that $\theta[\bx_i] \in R_i$. $Q(D)$ is defined as the set of all answers.}

A \emph{Union of Conjunctive Queries} $\varphi = \bigcup_{i \in \{1, \dots, \ell\}} \varphi_i$ is a set of CQs where head$(\varphi_{i_1}) = $ head$(\varphi_{i_2})$ for all $1 \leq i_1, i_2 \leq \ell$. Semantically, $\varphi(D) = \bigcup_{i \in \{1, \dots, \ell\}} \varphi_i(D)$. A UCQ is said to be full if each $\varphi_i$ is full.

\smallskip
\introparagraph{Natural Joins}
If a CQ is full, has no constants, and no repeated variables in the same atom, then we
say it is a {\em natural join query}. For instance, the 3-path query
$Q(x,y,z,w) = R(x,y), S(y,z), T(z,w)$ is a natural join query.
A natural join can be represented equivalently as a {\em hypergraph} 
$\mathcal{H}_Q = (\nodes_Q, \edges_Q)$, where $\nodes_Q$ is the set of variables, and
for each hyperedge $F \in \edges_Q$ there exists a relation $R_F$ with variables $F$. 
We will write the join as $\Join_{F \in \edges_Q} R_F$.
We denote the size of relation $R_F$  by $|R_F|$. Given two tuples $t_1$ and $t_2$ over a set of variables $\nodes_1$ and $\nodes_2$ where $\nodes_1 \cap \nodes_2 = \emptyset$, we will use $t_1 \circ t_2$ to denote the tuple formed over the variables $\nodes_1 \cup \nodes_2$. If $\nodes_1 \cap \nodes_2 \neq \emptyset$, then $t_1 \circ t_2$ will perform a join over the common variables.

\smallskip
\introparagraph{Join Size Bounds}
Let $\mathcal{H} = (\nodes, \edges)$ be a hypergraph, and $S \subseteq \nodes$.
A weight assignment $\bu = (u_F)_{F \in \edges}$ 
is called a {\em fractional edge cover} of $S$ if 
$(i)$ for every $F \in \edges, u_F \geq 0$  and $(ii)$ for every
$x \in S, \sum_{F:x \in F} u_F \geq 1$. 
The {\em fractional edge cover number} of $S$, denoted by
$\rho^*_{\mathcal{H}}(S)$ is the minimum of
$\sum_{F \in \edges} u_F$ over all fractional edge covers of $S$. 
We write $\rho^*(\mathcal{H}) = \rho^*_{\mathcal{H}}(\nodes)$.

In a celebrated result, Atserias, Grohe and Marx~\cite{AGM} proved that
for every fractional edge cover $\bu$ of $\nodes$,  the size 
of a natural join is bounded using the {\em AGM inequality}: $\label{eq:agm}
|\Join_{F \in \edges} R_F| \leq \prod_{F \in \edges} |R_F|^{u_F}
$
The above bound is constructive~\cite{skewstrikesback,ngo2012worst}: 
there exist worst-case algorithms that compute the join $\Join_{F \in \edges} R_F$ in 
time $O(\prod_{F \in \edges} |R_F|^{u_F})$ for every fractional edge cover $\bu$ of $\nodes$.

\smallskip
\introparagraph{Tree Decompositions}
Let $\mathcal{H} = (\nodes, \edges)$ be a hypergraph of a natural join query $Q$.
A {\em tree decomposition} of $\mathcal{H}$ is  a tuple 
$(\htree, (\bag_t)_{t \in V(\htree)})$ where $\htree$ is a tree, and 
every $\bag_t$ is a subset of $\nodes$, called the {\em bag} of $t$, such that 
	\begin{packed_enum}
		\item  
		each edge in $\edges$ is contained in some bag; and
		\item 
		for each variable $x \in \nodes$, the set of nodes $\{t \mid x \in \bag_t\}$ is connected in $\htree$.
	\end{packed_enum}
\noindent 
Given a rooted tree decomposition, we use $\prnt{t}$ to denote the (unique) parent of node $t \in V(\htree)$. 
Then, we define $\key{t} = \mB_t \cap \mB_{\prnt{t}}$ to be the common variables that occur in the bag $\mB_t$ and
its parent, and $\val{t} = \mB_t \setminus \key{t} $ the remaining variables of the bag.
We also use $\subtree{t}$ to denote the union of all bags in the subtree rooted at $t$ (including $\mB_t$).


The {\em fractional hypertree width} of a decomposition is 
defined as $\max_{t \in V(\htree)} \rho^*(\bag_t)$, where
$\rho^*(\bag_t)$ is the minimum fractional edge cover of the vertices in $\bag_t$.
The  fractional hypertree width of a query $Q$, denoted $\fhw(Q)$, is the minimum 
fractional hypertree width among all tree decompositions of its hypergraph.
We say that a query is {\em acyclic} if $\fhw(Q)=1$. \hlrone{If a query is acyclic, then there exists a tree decomposition such that the bags for the nodes of the decomposition correspond to the hyperedges $\edges$. Such a decomposition is known as a join tree.}
The {\em depth} of a rooted tree decomposition is  the largest distance over all root to leaf paths in $\htree$.


%

\smallskip
\introparagraph{Computational Model}
To measure the running time of our algorithms, we use the uniform-cost RAM 
model~\cite{hopcroft1975design}, where data values as well as pointers to
databases are of constant size. Throughout the article, all complexity results are 
with respect to data complexity (unless explicitly mentioned), where the query is assumed fixed. We will also use the set data structure that supports insertion and lookup of an element in constant time~\cite{cormen2022introduction}. In practice, hashing can only achieve amortized constant time for some of the operations. Therefore, all lookups, insertion times, and enumeration delays are amortized.

\subsection{Ranking Functions}

Consider a natural join query $Q$ and a database $D$. Our goal is to enumerate all
the tuples of $Q(D)$ according to an order that is specified by a {\em ranking function}.
In practice, this ordering could be specified, for instance, 
in the \oldtt{ORDER BY} clause of a \oldtt{SQL} query. 

Formally, we assume a total order $\succeq$ of the valuations $\theta$ over the variables
 of $Q$. The total order is induced by a ranking function $\rf$ that maps each valuation $\theta$ 
to a number $\rank{\theta} \in \mathbb{R}$. In particular, for two valuations $\theta_1, \theta_2$, 
we have $\theta_1 \succeq \theta_2$ if and only if $\rank{\theta_1} \geq \rank{\theta_2}$. Throughout the article, we will assume that $\rf$ is a computable function that takes times linear in the input size to the function. We present below two concrete examples of ranking functions.

\begin{exa} \label{ex:vertex}
For every constant $c \in \domain$, we associate a weight $w(c) \in \mathbb{R}$. 
Then, for each valuation $\theta$, we can define
$\rf(\theta) := \sum_{x \in \nodes} w(\theta(x)).$
This ranking function sums the weights of each value in the tuple. 
 \end{exa}

\begin{exa} \label{ex:edge}
For every input tuple $t \in R_F$, we associate a weight $w_F(t) \in \mathbb{R}$.
Then, for each valuation $\theta$, we can define
$\rank{\theta} = \sum_{F \in \edges} w_F(\theta[x_F])$
where $x_F$ is the set of variables in $F$. In this case, the ranking function sums the
weights of each contributing input tuple to the output tuple $t$ (we can extend the ranking function to all valuations
by associating a weight of 0 to tuples that are not contained in a relation).
\end{exa}

\smallskip
\introparagraph{Decomposable Rankings}
As we will see later, not all ranking functions are amenable to efficient evaluation. Intuitively, an arbitrary ranking
function will require that we look across all tuples to even find the smallest or largest element. We next
present several restrictions which are satisfied by ranking functions seen in practical settings.

\begin{defi}[Decomposable Ranking] \label{def:decomp}
Let $\rf$ be a ranking function over $\nodes$ and $S \subseteq \nodes$. We will use $\varphi$ to denote valuations
over the set of variables $\nodes \setminus S$. We say that $\rf$ is {\em $S$-decomposable} 
if there exists a total order for all valuations over $S$, such that for any two valuations $\theta_1, \theta_2$ over $S$ we have:
\hlrone{
$$ \theta_1 \succeq \theta_2 \Rightarrow 
\begin{cases}
    \forall \varphi, \rank{\varphi \circ \theta_1} =  \rank{\varphi \circ \theta_2} \\
    \text{or} \\
    \forall \varphi, \rank{\varphi \circ \theta_1} >  \rank{\varphi \circ \theta_2}
\end{cases}
$$}
\end{defi}

We say that a ranking function is {\em totally decomposable} if it is $S$-decomposable for every
subset $S \subseteq \nodes$, and that it is {\em coordinate decomposable} if it is $S$-decomposable
for any singleton set. Additionally, we say that it is {\em edge decomposable} for a query $Q$ if it is
$S$-decomposable for every set $S$ that is a hyperedge in the query hypergraph.
We point out here that totally decomposable functions are equivalent to monotonic orders as defined in~\cite{kimelfeld2006incrementally}.

\begin{exa}
The ranking function $\rank{\theta} = \sum_{x \in \nodes} w(\theta(x))$ defined in Example~\ref{ex:vertex} is 
totally decomposable, and hence also coordinate decomposable. Indeed, pick any set $S \subseteq \nodes$. 
We construct a total order on valuations $\theta$ over $S$ by using the value $\sum_{x \in S} w(\theta(x))$.
Now, consider valuations $\theta_1, \theta_2$ over $S$ such that
 $\sum_{x \in S} w(\theta_1(x)) \geq \sum_{x \in S} w(\theta_2(x))$.
Then, for any valuation $\varphi$ over $\nodes \setminus S$, \hlrone{if $\sum_{x \in S} w(\theta_1(x)) = \sum_{x \in S} w(\theta_2(x))$, we have:
\begin{align*}
\rank{\varphi \circ \theta_1}  \!=\!\! \sum_{x \in \nodes \setminus S} w(\varphi(x)) + \sum_{x \in S} w(\theta_1(x))
 = \!\sum_{x \in \nodes \setminus S} w(\varphi(x)) + \sum_{x \in S} w(\theta_2(x))
 = \rank{\varphi \circ \theta_2} 
\end{align*} 
Similarly, if $\sum_{x \in S} w(\theta_1(x)) > \sum_{x \in S} w(\theta_2(x))$, we have:
\begin{align*}
\rank{\varphi \circ \theta_1}  \!=\!\! \sum_{x \in \nodes \setminus S} w(\varphi(x)) + \sum_{x \in S} w(\theta_1(x))
 > \!\sum_{x \in \nodes \setminus S} w(\varphi(x)) + \sum_{x \in S} w(\theta_2(x))
 = \rank{\varphi \circ \theta_2} 
\end{align*} }

Next, we construct a function that is coordinate-decomposable but it is not totally decomposable. Consider the query 
$$Q(x_1 \dots, x_d, y_1, \dots, y_d) = R(x_1, \dots, x_d), S(y_1, \dots, y_d)$$ 
where $\domain = \{-1,1\}$, and define  $\rf(\theta) := \sum_{i=1}^d \theta(x_i) \cdot \theta(y_i)$. This ranking function corresponds to taking the inner product of the input tuples if viewed as vectors. The total order for $\domain$ is $-1 \prec 1$. It can be shown that for $d=2$, the function is not $\{x_1,x_2\}$-decomposable. \hlrone{For instance, if we define $\theta_1(x_1, x_2) = (1,-1) \succeq \theta_2(x_2, x_2) = (1,1)$, then for $\varphi(y_1, y_2) = (-1,-1)$ we get $\rf(\theta_1 \circ \varphi) = \rf(1,-1,-1,-1) = 1 \cdot (-1) + (-1) \cdot (-1) = 0 > \rf(\theta_2 \circ \varphi) = \rf(1,1,-1,-1) = 1 \cdot(-1) + 1 \cdot (-1) = -2 $ but if we define $\varphi = (1,1)$, then we get $\rf(\theta_1 \circ \varphi) = \rf(1,-1,1,1) = 1 \cdot 1 + 1 \cdot (-1) = 0 < \rf(\theta_2 \circ \varphi) = \rf(1,1,1,1) = 1 \cdot 1 + 1 \cdot 1$ = 2.} This demonstrates that the ranking function is not independent of valuations over $\{y_1, y_2\}$ and thus, the function does not satisfy the definition of decomposability.

\end{exa} 

\begin{defi}
Let $\rf$ be a ranking function over a set of variables $\nodes$, and $S,T \subseteq \nodes$
such that $S \cap T = \emptyset$.
We say that $\rf$ is {\em $T$-decomposable conditioned on $S$} 
if for every valuation $\theta$ over $S$, the function
$\rf_{\theta}(\varphi) := \rank{\theta \circ \varphi}$ defined over $\nodes \setminus S$ is $T$-decomposable.
\end{defi}
\noindent 
The next lemma connects the notion of conditioned decomposability with decomposability.

\begin{prop} \label{lem:conditioned}
Let $\mrf$ be a ranking function over a set of variables $\nodes$, and $T \subseteq \nodes$.
If $\mrf$ is $T$-decomposable, then it is also $T$-decomposable conditioned on $S$ for any $S \subseteq \nodes \setminus T$.
\end{prop}
\begin{proof}
	We need to show that for every valuation $\pi$ over $S$, $\rf(\pi \circ \Phi \circ \theta)$ is $T$-decomposable where $\Phi$ is defined over $U = \nodes \setminus (S \cup T)$ and $\theta$ is defined over $T$.  We use the same total order for $\theta$ as used for $T$-decomposability. Let $\theta_1 \succeq \theta_2$, and consider any valuation $\Phi$ over $U$. 
	Define the valuation $\varphi$ over $\nodes \setminus T$ such that $\varphi[S] = \pi$ and $\varphi[U] = \Phi$. Then,
	\hlrone{\begin{align*} 
	\theta_1 \succeq \theta_2 & \Rightarrow \begin{cases}
    \forall \varphi, \rank{\varphi \circ \theta_1} =  \rank{\varphi \circ \theta_2} \\
    \text{or} \\
    \forall \varphi, \rank{\varphi \circ \theta_1} >  \rank{\varphi \circ \theta_2}
    \end{cases} \\
	& \Leftrightarrow 
    \begin{cases}
    \forall \varphi, \rf(\varphi[S] \circ \varphi[U] \circ \theta_1) = \rf(\varphi[S] \circ \varphi[U] \circ \theta_2) \\
    \text{or} \\
    \forall \varphi, \rf(\varphi[S] \circ \varphi[U] \circ \theta_1) > \rf(\varphi[S] \circ \varphi[U] \circ \theta_2)
    \end{cases}
    \\
	& \Leftrightarrow 
    \begin{cases}
    \forall \varphi, \rf(\pi \circ \Phi \circ \theta_1) = \rf(\pi \circ \Phi \circ \theta_2) \\
    \text{or} \\
    \forall \varphi, \rf(\pi \circ \Phi \circ \theta_1) > \rf(\pi \circ \Phi \circ \theta_2)
    \end{cases}
	\end{align*}}
	Step $1$ follows from the definition of $T$-decomposable. Step $2$ and $3$ compute the restriction of $\varphi$ to $S$ and $U$.
\end{proof}

\noindent 
It is also easy to check that if a function is $(S\cup T)$-decomposable, then it is also $T$-decomposable
conditioned on $S$. 

\begin{defi}[Compatible Ranking]
Let $\htree$ be a rooted tree decomposition of hypergraph $\mathcal{H}$ of a natural join query. We say that a ranking function is {\em compatible
with $\htree$} if for every node $t$ it is $(\subtree{t} \setminus \key{t})$-decomposable conditioned on $\key{t}$.
\end{defi}

\begin{exa} \label{ex:decomposable}
Consider the join query $Q(x,y,z) = R(x,y), S(y,z)$, and the ranking function from Example~\ref{ex:edge},
 $\rf(\theta) = w_R(\theta(x),\theta(y)) + w_S(\theta(y), \theta(z))$. This function is not $\{z\}$-decomposable,
 but it is $\{z\}$-decomposable conditioned on $\{y\}$.
 
 Consider a decomposition of the hypergraph of $Q$ that has two nodes: the root node $r$ with $\mB_r = \{x,y\}$,
 and its child $t$ with $\mB_t = \{y,z\}$. Since $\subtree{t} = \{y,z\}$ and $\key{t} = \{y\}$, the condition of
 compatibility holds for node $t$. Similarly, for the root node $\subtree{t} = \{x,y,z\}$ and $\key{t} = \{\}$, hence the
 condition is trivially true as well. Thus, the ranking function is compatible with the decomposition. 
\end{exa}


\subsection{Problem Parameters}

Given a natural join query $Q$  and a database $D$, we want to enumerate
the tuples of $Q(D)$ according to the order  specified by $\rf$.
We will study this problem in the enumeration framework similar to that of~\cite{Segoufin15}, 
where an algorithm can be decomposed into two phases:

\begin{itemize}
	\item a {\bf preprocessing phase} that takes time $T_p$ and computes a data structure of size $S_p$,
	\item an {\bf enumeration phase} that outputs $Q(D)$ with no repetitions. The enumeration phase has full access to any data structures constructed in the preprocessing phase and can also use additional space of size $S_e$.
	The {\em delay} $\delta$ is defined as the maximum time to output any two consecutive tuples (and also the time to output the first tuple, and the time to notify that the enumeration has completed).  
\end{itemize}
\noindent 
It is straightforward to perform ranked enumeration for any ranking function by computing $Q(D)$, storing the tuples in an ordered list, and finally enumerating by scanning the ordered list with constant delay. This simple strategy implies the following result.

\begin{prop}\label{prop:simple:enum}
Let $Q$ be a natural join query with hypergraph $\mH = (\nodes, \edges)$. Let $\htree$ be a tree decomposition with fractional hypertree-width \mfhw, and $\mrf$ be a ranking function. Then, for any input database $D$, we can preprocess $D$ in time $ T_p = {O}(\log|D| \cdot |D|^{\fhw} + |Q(D)|)$ and space $S_p = O(|Q(D)|)$,
	such that for any $k$, we can enumerate the top-$k$ results of $Q(D)$ with
	$\text{delay } \delta = {O}(1)$ and $\text{space } S_e = O(1)$
\end{prop}
\noindent 
The drawback of Proposition~\ref{prop:simple:enum} is that the user will have to wait ${\Omega}(|Q(D)| \cdot \log |Q(D)|)$ time to even obtain the first tuple in the output. Moreover, even when we are interested in a few tuples, the whole output result will have to be materialized. Instead, we want to design algorithms that minimize the preprocessing time and space, while guaranteeing a small delay $\delta$. Interestingly, as we will see in~\autoref{sec:lowerbound}, the above result is essentially the best we can do if the ranking function is completely arbitrary; thus, we need to consider reasonable restrictions of $\rf$. 

To see what it is possible to achieve in this framework, it will be useful to keep in mind what we can do in the case where there is no ordering of the output.

\begin{thm}[due to \cite{DBLP:journals/tods/OlteanuZ15}]\label{thm:no:rank}
	Let $Q$ be a natural join query with hypergraph $\mH = (\nodes, \edges)$. Let $\htree$ be a tree decomposition with fractional hypertree-width \mfhw. Then, for any input database $D$, we can pre-process $D$ in time $T_p = O(|D|^{\mfhw})$ and space $ S_p = O(|D|^{\mfhw})$
	such that we can enumerate the results of $Q(D)$ with $\text{delay } \delta = {O}(1) $ and $\text{space } S_e = O(1)$
\end{thm}
\noindent 
For acyclic queries, $\fhw = 1$, and hence the preprocessing phase takes only linear time and space in the size of the input.

\section{Main Result}
\label{sec:proof}

In this section, we present our first main result. 


\begin{thm}[Main Theorem] \label{thm:main}
Let $Q$ be a natural join query with hypergraph $\mH = (\nodes, \edges)$. Let $\htree$ be a fixed tree decomposition with fractional hypertree-width \mfhw, and $\mrf$ be a ranking function that is compatible with $\htree$.
Then, for any database $D$, we can preprocess $D$ with
	$$
		 T_p = {O}(|D|^{\mfhw}) \quad \quad S_p = {O}(|D|^{\mfhw})
	$$
	such that for any $k$, we can enumerate the top-$k$ tuples of $Q(D)$ with
	$$
		 \text{delay } \delta = {O}(\log |D|) \quad \quad
		\text{space } S_e = O(\min \{k, \vert Q(D) \vert\})
	$$
	\end{thm}
\vspace{0.5\baselineskip}
\noindent 
In the above theorem, the preprocessing step is independent of the value of $k$: we perform exactly the same preprocessing if the user only wants to obtain the first tuple, or all tuples in the result. However, if the user decides to stop after having obtained the first $k$ results, the space used during enumeration will be bound by $O(k)$.
We should also note that all of our algorithms work in the case where the ordering of the tuples/valuations is instead expressed through a \texttt{comparable} function that, given two valuations, returns the larger.

It is instructive to compare \autoref{thm:main} with \autoref{thm:no:rank}, where no ranking is used when enumerating the results. There are two major differences. First, the delay $\delta$ has an additional logarithmic factor. As we will discuss later in~\autoref{sec:lowerbound}, this logarithmic factor is a result of doing ranked enumeration, and it is most likely unavoidable. The second difference is that the space $S_e$ used during enumeration blows up from constant $O(1)$ to $O(|Q(D)|)$ in the worst case (when all results are enumerated). \hlrone{Let us now define some notation that we will use in the section. Given a set $U$, $\langle U, \oplus \rangle$ is said to be a {\em commutative monoid} if $\oplus$ is a binary operator that is commutative, associative, and has an identity element in $U$. We say that the operator $\oplus$ is strictly monotone if $a = b$ implies that $a \oplus c = b \oplus c$, and if $a > b$ then $a \oplus c > b \oplus c$ for every $a,b,c \in U$. $\langle \mathbb{R}, + \rangle$ and $\langle \mathbb{N}^+, * \rangle$ are examples of commutative monoids where $\oplus$ is strictly monotone.}

In the remainder of this section, we will present a few applications of \autoref{thm:main}, and then prove the theorem.

\subsection{Applications}
\label{sec:app}

We show here how to apply \autoref{thm:main} to obtain algorithms for different ranking functions.

\smallskip
\introparagraph{Vertex-Based Ranking} A vertex-based ranking function over $\nodes$ is of the form: $ \rf(\theta) := \bigoplus_{x \in \nodes} f_x(\theta(x)) $
where $f_x$ maps values from $\domain$ to some set $U \subseteq \mathbb{R}$ and 
$\langle U, \oplus \rangle$ forms a {commutative monoid}. 

\begin{lem}\label{lem:vb}
Let $\mrf$ be a \hlrone{strictly} monotone vertex-based ranking function over $\nodes$. Then, $\mrf$ is totally decomposable,
and hence compatible with any tree decomposition of a hypergraph with vertices $\nodes$.
\end{lem}
\begin{proof}
	
	Pick any set $S \subseteq \nodes$ and let
	$\theta^\star$ be the valuation over $\nodes \setminus S$ such that for every $x$, $f_x(\theta^\star(x)) = e$,
	where $e$ is the identity element of the monoid. 
	\hlrone{We will define a total order over $S$ in the following way. Since $f_x$ maps values from $\domain$ to $U$, it holds that  $ \rank{\theta^\star \circ \theta_1}$ and $\rank{\theta^\star \circ \theta_2}$ are comparable for any two valuations $\theta_1, \theta_2$ that are defined over $S$.
	Therefore, if $ \rank{\theta^\star \circ \theta_1} \geq  \rank{\theta^\star \circ \theta_2}$, then $\theta_1 \succeq \theta_2$ and vice-versa. This establishes a total order.

	\eat{Suppose that $ \rank{\theta^\star \circ \theta_1} \geq  \rank{\theta^\star \circ \theta_2}$ for
	valuations $\theta_1, \theta_2$ over $S$ establishing a total order $\theta_1 \succeq \theta_2$.}
	\vspace{0.5\baselineskip}
	\noindent 
	Therefore, if $\theta_1 \succeq \theta_2$, it holds that
	$\oplus_{x \in S} f_x(\theta_1(x)) = \oplus_{x \in S} f_x(\theta_2(x))$ or $\oplus_{x \in S} f_x(\theta_1(x)) > \oplus_{x \in S} f_x(\theta_2(x))$.

	\vspace{0.5\baselineskip}
	If $\oplus_{x \in S} f_x(\theta_1(x)) = \oplus_{x \in S} f_x(\theta_2(x))$, for any valuation $\theta$ over $\nodes \setminus S$ we have:
	\begin{align*}
	\rank{\theta \circ \theta_1} & =  \oplus_{x \in \nodes \setminus S} f_x(\theta(x))  \bigoplus  \oplus_{x \in S} f_x(\theta_1(x))  \\
	& = \oplus_{x \in \nodes \setminus S} f_x(\theta(x))  \bigoplus  \oplus_{x \in S} f_x(\theta_2(x))  \\
	& = \rank{\theta \circ \theta_2}
	\end{align*} 

    Similarly, if $\oplus_{x \in S} f_x(\theta_1(x)) > \oplus_{x \in S} f_x(\theta_2(x))$, for any valuation $\theta$ over $\nodes \setminus S$ we have:
	\begin{align*}
	\rank{\theta \circ \theta_1} & =  \oplus_{x \in \nodes \setminus S} f_x(\theta(x))  \bigoplus  \oplus_{x \in S} f_x(\theta_1(x))  \\
	& > \oplus_{x \in \nodes \setminus S} f_x(\theta(x))  \bigoplus  \oplus_{x \in S} f_x(\theta_2(x))  \\
	& = \rank{\theta \circ \theta_2}
	\end{align*} 
	The inequalities holds because of the strict monotonicity of the binary operator.}
\end{proof}

\introparagraph{Tuple-Based Ranking} 
Given a query hypergraph $\mH$, \hlrone{suppose we assign for every valuation} $\theta$ over
the variables $x_F$ of relation $R_F$ a weight $w_F(\theta) \in U \subseteq \mathbb{R}$. Then, \hlrone{a tuple-based ranking function takes the}
following form: $ \rf(\theta) := \bigoplus_{F \in \edges} w_F(\theta[x_F]) $
where $\langle U, \oplus \rangle$ forms a {commutative monoid}. In other words, a tuple-based ranking function
assigns a weight to each input tuple, and then combines the weights through $\oplus$.

\begin{lem} \label{lem:tb}
Let $\mrf$ be a \hlrone{strictly} monotone tuple-based ranking function over \hlrone{the hypergraph $\mH = (\nodes, \edges)$}. Then, $\mrf$
is compatible with any tree decomposition of the hypergraph.
\end{lem}
\begin{proof}
	Pick some node $t$ in the decomposition, and fix a valuation $\theta_0$ over $\key{t}$.
	Let $E \subseteq \edges$ be the hyperedges that correspond to bags in the subtree rooted at $t$,
	and $\bar{E}$ the remaining hyperedges.
	Let $\theta^\star$ be the valuation over $\nodes \setminus \subtree{t}$ such that for every $F \in \bar{E}$
	we have $w_F((\theta_0 \circ \theta^\star)[x_F]) =e$, where $e$ is the identity element. Notice that the
	latter is well-defined, since the hyperedges in $\bar{E}$ can not contain any variables in $\subtree{t} \setminus \key{t}$.
	
	\hlrone{We will define a total order over $\subtree{t} \setminus \key{t}$ in the following way. Since $w_F$ maps values to $U$, it holds that  $ \rank{\theta_0 \circ \theta^\star \circ \theta_1}$ and $\rank{\theta_0 \circ \theta^\star \circ \theta_2}$ are comparable for any two valuations $\theta_1, \theta_2$ over $S$. Therefore, if $\rank{\theta_0 \circ \theta^\star \circ \theta_1} \geq  \rank{\theta_0 \circ \theta^\star \circ \theta_2}$, then $\theta_1 \succeq \theta_2$ and vice-versa. This establishes a total order.
	Given $\theta_1$ and $\theta_2$ such that $\theta_1 \succeq \theta_2$, we have that either 
	$\oplus_{F \in E} w_F((\theta_0 \circ \theta_1)[x_F]) = \oplus_{F \in E} w_F((\theta_0 \circ \theta_2)[x_F])$ or $\oplus_{F \in E} w_F((\theta_0 \circ \theta_1)[x_F]) > \oplus_{F \in E} w_F((\theta_0 \circ \theta_2)[x_F])$.
    
	If the former is true, then for any valuation $\theta$ over $\nodes \setminus \subtree{t}$, we have:
	\begin{align*}
		\rf(\theta_0 & \circ \theta \circ \theta_1) = \\
		& =  \oplus_{F \in E} w_F((\theta_0 \circ \theta_1)[x_F])  \bigoplus  \oplus_{F \in \bar{E}} w_F((\theta \circ \theta_0)[x_F])   \\
		& = \oplus_{F \in E} w_F((\theta_0 \circ \theta_2)[x_F])  \bigoplus  \oplus_{F \in \bar{E}} w_F((\theta \circ \theta_0)[x_F])   \\
		& = \rf(\theta_0  \circ \theta \circ \theta_2) 
	\end{align*} 
    For the latter, we have:
    \begin{align*}
		\rf(\theta_0 & \circ \theta \circ \theta_1) = \\
		& =  \oplus_{F \in E} w_F((\theta_0 \circ \theta_1)[x_F])  \bigoplus  \oplus_{F \in \bar{E}} w_F((\theta \circ \theta_0)[x_F])   \\
		& > \oplus_{F \in E} w_F((\theta_0 \circ \theta_2)[x_F])  \bigoplus  \oplus_{F \in \bar{E}} w_F((\theta \circ \theta_0)[x_F])   \\
		& = \rf(\theta_0  \circ \theta \circ \theta_2) 
	\end{align*} 
    
	The inequality hold because of the strict monotonicity of the binary operator.  }
\end{proof}
\noindent 
Since both monotone tuple-based and vertex-based ranking functions are compatible with any tree decomposition
we choose, the following result is immediate.

\begin{prop} \label{prop:rankings}
Let $Q$ be a natural join query with optimal fractional hypertree-width $\mfhw$. Let $\mrf$ be a ranking function that can be 
either $(i)$ strictly monotone vertex-based, $(ii)$ strictly monotone tuple-based.
Then, for any input $D$, we can pre-process $D$ in time $ T_p = {O}(|D|^{\mfhw})$ and space $ S_p = {O}(|D|^{\mfhw})$	such that for any $k$, we can enumerate the top-$k$ results of $Q(D)$ with $ \delta = O(\log|D|) $ and  $ S_e = O(\min \{k, \vert Q(D) \vert\})$
\end{prop}

For instance, if the query is acyclic, hence $\fhw=1$, the above theorem gives an algorithm with linear preprocessing time $O(|D|)$ and $O(\log |D|)$ delay.

\smallskip
\introparagraph{Lexicographic Ranking} A typical ordering of the output valuations is according to a {\em lexicographic order}.
In this case, each $\domain(x)$ is equipped with a total order. If $\nodes = \{x_1, \dots, x_k\}$, a lexicographic order
$\langle x_{i_1}, \dots, x_{i_\ell} \rangle$ for $\ell \leq k$ means that two valuations $\theta_1, \theta_2$ are first ranked on $x_{i_1}$, and if they have the same rank on $x_{i_1}$, then they are ranked on $x_{i_2}$, and so on. This ordering can be naturally encoded by first taking a function $f_x: \domain(x) \rightarrow \mathbb{R}$ that captures the total order for variable $x$, and then defining $\rf(\theta) := \sum_x w_x f_x(\theta(x))$, where $w_x$ are appropriately chosen values. \hlrone{Suppose the size of the domain is $n$. Then, the weight for variable $x_i$ can we set as $w_x = n^{i}$. For example, if $\domain(x) = \{0, 1, \dots, 9\}$, then fixing $w_{x_i} = 10^i$ allows us to compare any two tuples by looking at their score computed using the ranking function. Since this example ranking} function is a monotone vertex-based ranking, \autoref{prop:rankings} applies here as well. 

We should note here that lexicographic ordering has been previously considered in the context of factorized databases.

\begin{prop}[due to ~\cite{DBLP:journals/tods/OlteanuZ15, bakibayev2012fdb}] \label{thm:factorized:lexicographic}
Let $Q$ be a natural join query with hypergraph $\mH = (\nodes, \edges)$, and 
$\langle x_{i_1}, \dots, x_{i_\ell} \rangle$  a lexicographic ordering of the variables in $\nodes$.

\vspace{0.5\baselineskip}
\noindent 
Let $\htree$ be a tree decomposition with fractional hypertree-width \fhwob\ such that 
$\langle x_{i_1}, \dots, x_{i_\ell} \rangle$ forms a prefix in the topological ordering of the variables in the decomposition. 
Then, for any input database $D$, we can pre-process $D$ with $T_p = O(|D|^{\fhwob})$ and  $S_p = O(|D|^{\fhwob})$
such that results of $Q(D)$ can be enumerated with delay $\delta = {O}(1)$ and space $S_e = O(1)$.	
\end{prop}

\noindent 
In other words, if the lexicographic order ``agrees" with the tree decomposition (in the sense that whenever $x_{i}$ is before $x_j$ in the lexicographic order, $x_j$ can never be in a bag higher than the bag where $x_i$ is), then it is possible to get an even better result than \autoref{thm:main}, by achieving constant delay $O(1)$, and constant space $S_e$.
However, given a tree decomposition, \autoref{thm:main} applies for any lexicographic ordering - in contrast to \autoref{thm:factorized:lexicographic}. 
As an example, consider the join query $Q(x,y,z) = R(x,y),S(y,z)$ and the lexicographic ordering $\langle z, x, y \rangle$. 
Since $\fhw = 1$, our result implies that we can achieve $O(|D|)$ time preprocessing with delay ${O}(\log |D|)$.
On the other hand, the optimal width of a tree decomposition that agrees with $\langle z, x, y \rangle$ is $\fhwob = 2$; 
hence, \autoref{thm:factorized:lexicographic} implies $O(|D|^2)$ preprocessing time and space. 
Thus, variable orderings in a decomposition fail to capture the additional challenge of user chosen lexicographic orderings. It is also not clear whether further restrictions on variable orderings in \autoref{thm:factorized:lexicographic} are sufficient to capture ordered enumeration for other ranking functions (such as sum).

\smallskip
\introparagraph{Bounded Ranking} A ranking function is {\em $c$-bounded} if there exists a subset $S \subseteq \nodes$ of
size $|S| =c$, such that the value of $\rf$ depends only on the variables from $S$. A $c$-bounded ranking is related to $c$-determined ranking functions~\cite{kimelfeld2006incrementally}: $c$-determined implies $c$-bounded, but not vice versa.
For $c$-bounded ranking functions, we can show the following result:

\begin{prop}  \label{prop:c}
Let $Q$ be a natural join query with optimal fractional hypertree-width $\mfhw$. If $\mrf$ is a $c$-bounded ranking function,
then for any input $D$, we can pre-process $D$ in time $ T_p = {O}(|D|^{\mfhw+c})$ and space $ S_p = {O}(|D|^{\mfhw+c})	$ such that for any $k$, we can enumerate the top-$k$ results of $Q(D)$ with	$\delta = O(\log |D|)$ and $ S_e = O(\min \{k, \vert Q(D) \vert\})$
\end{prop}
\begin{proof}
	Let $\htree$ by the optimal decomposition of $Q$ with fractional hypertree-width $\fhw$. We create a new decomposition $\htree'$ by simply adding the variables $S$ that determine the ranking functions in all the bags of $\htree$. By doing this, the width of the decomposition will grow by at most an additive factor of $c$. To complete the proof, we need to show that $\rf$ is compatible with the new decomposition. 
	
	Indeed, for any node in $\htree'$ (with the exception of the root node) we have that $S \subseteq \key{t}$. Hence, if we fix a valuation over $\key{t}$, the ranking function will output exactly the same score, independent of what values the other variables take. 
\end{proof}

\subsection{The Algorithm for the Main Theorem}

At a high level, each node $t$ in the tree decomposition will materialize, in an incremental fashion, all valuations over
$\subtree{t}$ that satisfy the join query corresponding to the subtree rooted at $t$. 
We will not store explicitly each valuation $\theta$ over $\subtree{t}$ at every node $t$, but instead we use a simple 
recursive structure $c(v)$ that we call a {\em cell}.
If $t$ is a leaf, then $c(\theta) = \langle \theta, [], \bot \rangle$, where $\bot$ is used to denote a null pointer.
Otherwise, suppose that $t$ has $n$ children $t_1, \dots, t_n$.
Then, $c(\theta) = \langle \theta[\mB_t], [p_1, \dots, p_n], q \rangle$, where 
$p_i$ is a pointer to the cell $c(\theta[\subtree{t_i}])$ stored at node $t_i$,
and $q$ is a pointer to a cell stored at node $t$ (intuitively representing the ``next" valuation in the order). \hlrone{We will use the notation $c(\theta).\texttt{FIRST()}$ and $ c(\theta).\texttt{MID()}$ to refer to the first and middle values of the triple denoted by the cell.} Given a cell $c(\theta)$, Algorithm~\ref{algo:reconstruct} shows how to reconstruct the full tuple in constant time (dependent only on the query) by traversing the subtree rooted at node $t$.

\begin{algorithm}
	\SetCommentSty{textsf}
	\DontPrintSemicolon 
	\SetKwFunction{fullreducer}{\textsc{Materialize}}
	\SetKwFunction{initializepq}{\textsc{InitializePQ}}
	\SetKwFunction{recurse}{\textsc{Recurse}}
	\SetKwFunction{proc}{\textsf{eval}}
	\SetKwFunction{popfront}{\textsc{popfront()}}
	\SetKwFunction{pushback}{\textsc{pushback}}
	\SetKwFunction{first}{\textsc{first()}}
	\SetKwFunction{mid}{\textsc{mid()}}
	\SetKwFunction{last}{\textsc{last()}}		
	\SetKwFunction{pop}{\textsc{pop()}}
	\SetKwData{pointer}{\textsf{pointer}}
	\SetKwFunction{preprocess}{\textsc{preprocess}}	
	\SetKwData{temp}{\textsf{temp}}
	\SetKwFunction{fillout}{\textsc{FillOUT}}
	\SetKwInOut{Input}{\textsc{input}}
	\SetKwInOut{Global}{\textsc{global variables}}
	\SetKwInOut{Output}{\textsc{output}}
	\SetKwProg{myproc}{\textsc{procedure}}{}{}
    \hlrone{\Input{Cell $c$ for node $t$}
	\Output{Tuple $\out(c)$}
	$L \leftarrow [c]$ \tcc*{List $L$ supports popping from the front and pushing to the back.}
	\While{$L$ is not empty}{
		$s \leftarrow L.\popfront$ \;
		$\gamma_s \leftarrow s.\first$\;
		\ForEach{$p \in s.\mid$}{
			$L.\pushback(*p)$ \tcc*{Push the cell at address $p$ at the end of $L$.}
		}
	}
	$\theta \leftarrow \Join_{s \in \subtree{t}} \gamma_s$\;
	\textbf{return} $\theta$
	\caption{Constructing tuple represented by a cell $c$}}
	\label{algo:reconstruct}
\end{algorithm}

\hlrone{Given a cell $c$, we will refer to the tuple constructed by Algorithm~\ref{algo:reconstruct} as $\out(c)$.} Next, each node $t$ maintains one hash map $\qQ_t$, which
maps each valuation $u$ over $\key{t}$ to a {\em priority queue} $\qQ_t[u]$.
The elements of $\qQ_t[u]$ is a \hlrone{pair $\langle \mathit{score}, c(\theta) \rangle$, where $\theta$ is a valuation over $\subtree{t}$
such that $u = \theta[\key{t}]$ and \textit{score} is the value assigned to the cell by the ranking function and it is used by the priority queue.} The priority queues will be the data structure that performs the
comparison and ordering between different tuples. We will use an implementation of a priority queue (e.g., a Fibonacci heap~\cite{CLRS}) with the following properties: $(i)$ we can insert an element in constant time $O(1)$, $(ii)$ we can obtain the min element (top) in time $O(1)$, and $(iii)$ we can delete the min element (pop) in time $O(\log n)$. 

Notice that it is not straightforward to rank the cells according to the valuations, since the ranking function is
defined over all variables $\nodes$. However, here we can use the fact that the ranking function is compatible
with the decomposition at hand. For each variable $x \in \nodes$, we designate some value from $\domain(x)$ as $v^\star(x)$. Given a fixed valuation $u$ over $\key{t}$, we will order the valuations $\theta$ 
over $\subtree{t}$ that agree with $u$ according to the score:
$ \rank{v_{t}^{\star} \circ \theta}$
where $v_{t}^\star = (v^\star(x_1), v^\star(x_2),\dots, v^\star(x_p))$ is a valuation over the $p$ variables $S = \nodes \setminus \subtree{t}$. The key intuition is that the compatibility of the ranking function with the decomposition implies that the ordering of the tuples implied by the cells in the priority queue $\qQ_t[u]$ will not change if we replace $v_{t}^\star$ with any other valuation. Thus, the comparator can use $v_{t}^\star$ to calculate the score which is used by the priority queue internally.
We next discuss the {\em preprocessing} and {\em enumeration} phase of the algorithm.

		\begin{algorithm}
			\SetCommentSty{textsf}
			\DontPrintSemicolon 
			\SetKwFunction{fullreducer}{\textsc{Materialize}}
			\SetKwFunction{initializepq}{\textsc{InitializePQ}}
			\SetKwFunction{recurse}{\textsc{Recurse}}
			\SetKwFunction{proc}{\textsf{eval}}
			\SetKwFunction{ins}{\textsc{insert}}
			\SetKwFunction{app}{\textsc{insert}}
			\SetKwFunction{top}{\textsc{top()}}
			\SetKwFunction{pop}{\textsc{pop()}}
			\SetKwData{pointer}{\textsf{pointer}}
			\SetKwFunction{preprocess}{\textsc{preprocess}}	
			\SetKwData{temp}{\textsf{temp}}
			\SetKwFunction{fillout}{\textsc{FillOUT}}
			\SetKwInOut{Input}{\textsc{input}}
			\SetKwInOut{Global}{\textsc{global variables}}
			\SetKwInOut{Output}{\textsc{output}}
			\SetKwProg{myproc}{\textsc{procedure}}{}{}
			 \hlrone{\Input{CQ $Q$, Tree decomposition $(\htree, \mB_{t \in V(\htree)})$, Database $D$, Ranking function $\rf$}}
             \hlrone{\Output{Initialized priority queues $\qQ_t$ and a set $H[t]$ for each node $t \in V(\htree)$}}
				\ForEach{$t \in V(\htree)$}{
					\textit{materialize the bag} $\mB_t$ 
                    \tcc*{$R_t$ denotes the materialized relation for node $t$}
				}
				\textit{full reducer pass on materialized bags in} $\htree$ \;
				
				\BlankLine
				\ForAll{$t \in V(\htree)$ in post-order traversal}{
                
                    \hlrone{$H[t] \gets $empty set}\;
					\ForEach{valuation $\theta$ in the relation corresponding to $t$}{
						$u \gets \theta[\key{t}]$\;
						\If{$\qQ_t[u]$ is $\texttt{NULL}$}{ $\qQ_t[u] \gets$ new priority queue}
						$\ell \gets []$ \tcc*{$\ell$ is a list of pointers}
						\ForEach{child $s$ of $t$}{ \label{line:loop}
							\label{line:topq} $\ell.\app( \qQ_s[\theta[\key{{s}}]].\top)$ 
						}
                        $c \gets \langle \theta, \ell, \bot \rangle \rangle$\;
						\hlrone{$H[t].\ins(\out{(c)})$}\;           $\hlrone{\qQ_t[u].\ins(\langle \rank{v_{t}^{\star} \circ \theta}, c)}$ \label{pre:insert}
					}
				}
			
			\caption{Preprocessing Phase}
			\label{algo:preprocessing}
	\end{algorithm}

		\begin{algorithm}
			\SetCommentSty{textsf}
			\DontPrintSemicolon 
			\SetKwFunction{fullreducer}{\textsc{Materialize}}
			\SetKwFunction{initializepq}{\textsc{InitializePQ}}
			\SetKwFunction{fillout}{\textsc{topdown}}
			\SetKwFunction{ins}{\textsc{insert}}
			\SetKwFunction{app}{\textsc{insert}}
			\SetKwFunction{top}{\textsc{top()}}
			\SetKwFunction{pop}{\textsc{pop()}}
			\SetKwFunction{enum}{\textsc{enum}}	
			\SetKwFunction{eoe}{\textsc{address\_of}}	
			\SetKwData{pointer}{\textsf{pointer}}
            \SetKwFunction{first}{\textsc{first()}}
        	\SetKwFunction{mid}{\textsc{mid()}}
        	\SetKwFunction{last}{\textsc{last()}}	
			\SetKwData{temp}{\textsf{temp}}
			\SetKwData{retval}{\textsf{retval}}
			\SetKwData{popvalue}{\textsf{popvalue}}
			\SetKwInOut{Input}{\textsc{input}}\SetKwInOut{Output}{\textsc{output}}
			\SetKwProg{myproc}{\textsc{procedure}}{}{}
			\hlrone{\Input{CQ $Q$, Tree decomposition $(\htree, \mB_{t \in V(\htree)})$, Database $D$, Ranking function $\rf$, Initialized priority queues $\qQ_t$ and a set $H[t]$ for each node $t \in V(\htree)$}}
             \hlrone{\Output{Enumerates $Q(D)$ sorted according to $\rf$}}
			\myproc{\enum{} \label{func:enum}}{
				\While{$\qQ_{r}[()]$ is not empty}{
					\textbf{print} $\hlrone{\out(\qQ_{r}[()].\top)}$ \;
					\fillout{$\qQ_{r}[()].\top, r$} \;
				}
			}
			\BlankLine
			\myproc{\fillout{$c, t$}}{
				/* $c = \langle \theta,  [p_1, \dots, p_k], \mnext \rangle$ */ \;
				$u \gets \theta[\key{t}]$ \tcc*{$\theta = c.\first$}
				\If{$\mnext == \bot$ \label{line:if}}{
					$b \leftarrow \qQ_t[u].\pop$ \label{tuple:pop}\;
					\ForEach{child $t_i$ of $t$ \label{for:1}}{
						$p_i' \gets $\fillout{$*p_i, t_i$} \tcc*{$p_i = c.\mid[i]$}
                        \If{$p_i' \neq \bot$}{
                        \hlrone{$c' \gets \langle  \theta,  [p_1, \dots, p_i', \dots p_k], \bot \rangle \rangle$\label{push:queue}\;
						\If{$\out{(c')} \not\in H[t]$ \label{dupli} \tcc*{avoiding addition of duplicate cells}}{ 
                        $H[t].\ins(\out{(c')})$\;}
                        $\hlrone{\qQ_t[u].\ins((\langle \rank{v_{t}^{\star} \circ \theta}, c')}$ \label{tuple:add}
						}
					}}
					\If {$t$ is not the root}{
						$\mnext \gets \eoe(\qQ_t[u].\top)$ \label{chaining} \tcc*{$\mnext$ is an alias for $c.\last$}
						
					}
				}
				\KwRet{$\mnext$} \;
			}
			\caption{Enumeration Phase} \label{algo:enumeration}
	\end{algorithm}

\smallskip
\introparagraph{Preprocessing}
Algorithm~\ref{algo:preprocessing} consists of two steps.
The first step works exactly as in the case where there is no ranking function: each bag $\mB_t$ is computed and materialized, and then we apply a full reducer pass to remove all tuples from the materialized bags that will not join in the final result. 
The second step initializes the hash map with the priority queues for every bag in the tree. We traverse the decomposition in a bottom up fashion (post-order traversal), and do the following. For a leaf node $t$, notice that the algorithm does not enter the loop in line~\ref{line:loop}, so each valuation $\theta$ over $\mB_t$ is added to the corresponding queue as the triple $\langle \theta, [], \bot \rangle$. For each non-leaf node $t$, we take each valuation $v$ over $\mB_t$ and form a valuation (in the form of a cell) over $\subtree{t}$ by using the valuations with the largest rank from its children (we do this by accessing the top of the corresponding queues in line~\ref{line:topq}). The cell is then added to the corresponding priority queue of the bag. Observe that the root node $r$ has only one priority queue, since $\key{r} = \{\}$.

\begin{exa} \label{ex:initial}
	As a running example, we consider the  natural join query $Q(x,y,z,p) =  R_1(x,y), R_2(y,z), R_3(z,p), R_4(z,u)$ where the ranking function is the sum of the weights of each input tuple. Consider the following instance $D$ and decomposition $\htree$ for our running example.
	
	\begin{minipage}[t]{0.3\linewidth}
		\centering
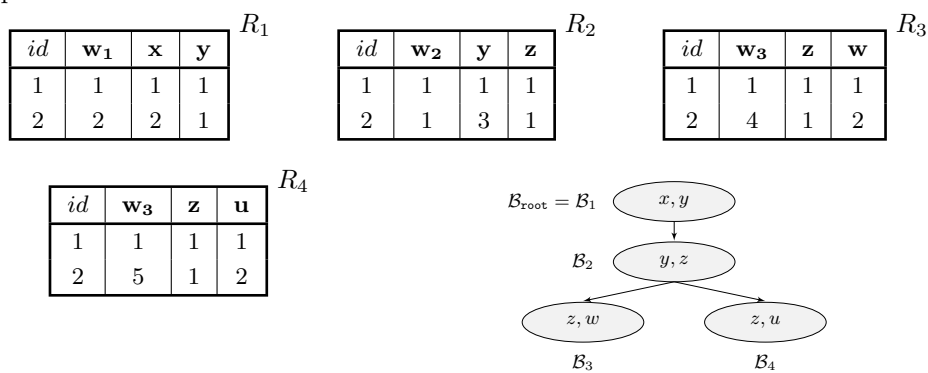
		\begin{tabular}[t]{ !{\vrule width1pt} c| c|c|c  !{\vrule width1pt} } 
			\Xhline{1pt}
			$id$ & $\mathbf{w_1}$ & $\mathbf{x}$ & $\mathbf{y}$ \\ 
			\Xhline{1pt}
			1 & 1 & 1 & 1 \\ 
			2& 2 & 2 & 1 \\
			\Xhline{1pt}
		\end{tabular}		
		\vspace{1em}
		$R_1$
	\end{minipage}
	\begin{minipage}[t]{0.3\linewidth}
		\centering
		\begin{tabular}[t]{ !{\vrule width1pt} c| c|c|c !{\vrule width1pt} } 
			\Xhline{1pt}
			$id$& $\mathbf{w_2}$ & $\mathbf{y}$ & $\mathbf{z}$ \\ 
			\Xhline{1pt}
			1 & 1 & 1 & 1 \\ 
			2 & 1 & 3 & 1 \\
			\Xhline{1pt}
		\end{tabular}		
		\vspace{1em}	
		$R_2$
	\end{minipage}
	\begin{minipage}[t]{0.3\linewidth}
		\centering
		\begin{tabular}[t]{ !{\vrule width1pt} c|c|c|c !{\vrule width1pt} } 
			\Xhline{1pt}
			$id$ & $\mathbf{w_3}$ & $\mathbf{z}$ & $\mathbf{p}$ \\ 
			\Xhline{1pt}
			1 & 1 & 1 & 1 \\ 
			2 & 4 & 1 & 2 \\
			\Xhline{1pt}
		\end{tabular}		
		\vspace{1em}
		$R_3$
	\end{minipage}
	\newline
	\begin{minipage}[t]{0.45\linewidth}
		\centering
		\begin{tabular}[t]{ !{\vrule width1pt} c|c|c|c !{\vrule width1pt} } 
			\Xhline{1pt}
			$id$ & $\mathbf{w_3}$ & $\mathbf{z}$ & $\mathbf{u}$ \\ 
			\Xhline{1pt}
			1 & 1 & 1 & 1 \\ 
			2 & 5 & 1 & 2 \\
			\Xhline{1pt}
		\end{tabular}		
		\vspace{1em}
		$R_4$
	\end{minipage}
	\begin{minipage}[t]{0.45\linewidth}
		\centering
		\vspace{-1.5em}
		\scalebox{.8}{\begin{tikzpicture}
			\tikzset{edge/.style = {->,> = latex'},
				vertex/.style={circle, thick, minimum size=7mm}}
			\def\x{0.25}
			
			\begin{scope}[fill opacity=1]
			\draw[fill=black!5] (0, 0) ellipse (1cm and 0.33cm) node {\small ${\color{black} x, y}$};
			\draw[fill=black!5] (0, -1) ellipse (1cm and 0.33cm) node {\small ${\color{black} y, z}$};
			\draw[fill=black!5] (-1.5, -2) ellipse (1cm and 0.33cm) node {\small ${\color{black} z, p}$};
			\draw[fill=black!5] (1.5, -2) ellipse (1cm and 0.33cm) node {\small ${\color{black} z, u}$};
			\draw[edge] (0, -0.33) -- (0,-.65);
			\draw[edge] (0, -1.33) -- (-1.5, -1.65);
			\draw[edge] (0, -1.33) -- (1.5, -1.65);
			
			\node[vertex]  at (-2, 0) {\small $\mB_{\texttt{root}} = \mB_1$};
			\node[vertex]  at (-1.5, -1) {\small $\mB_{2}$};				
			\node[vertex]  at (-1.5, -2.66) {\small $\mB_{3}$};				
			\node[vertex]  at (1.5, -2.66) {\small $\mB_{4}$};												
			\end{scope}	
			\end{tikzpicture}
		}
	\end{minipage}

\begin{figure}[htp!]
	\begin{subfigure}[t]{0.99\linewidth}
		\centering
		\scalebox{.99}{\begin{tikzpicture}
			\tikzset{edge/.style = {->,> = latex'},
				vertex/.style={circle, thick, minimum size=7mm}}
			\def\x{0.25}
			\def\y{-1}
			\begin{scope}[fill opacity=0.5, draw opacity = 0.5]
			\draw[fill=black!5] (0, 1) ellipse (1cm and 0.33cm) node {\small ${\color{black} x, y}$};
			\draw[fill=black!5] (0, -1) ellipse (1cm and 0.33cm) node {\small ${\color{black} y, z}$};
			\draw[fill=black!5] (-3, -2.25) ellipse (1cm and 0.33cm) node {\small ${\color{black} z, p}$};
			\draw[fill=black!5] (3, -2.25) ellipse (1cm and 0.33cm) node {\small ${\color{black} z, u}$};
			\draw[edge] (0, 0.66) -- (0,-.65);
			\draw[edge] (0, -1.33) -- (-3, -1.9);
			\draw[edge] (0, -1.33) -- (3, -1.9);
			
			\end{scope}	
			\tikzset{pblock/.style = {rectangle split, rectangle split parts=2,
					draw, rectangle split horizontal=false,inner ysep=0pt, line width=0.15em, text width=8em, align=center, rectangle split part fill = {white, white}}}
			
			\tikzset{qblock/.style = {rectangle split, rectangle split parts=1,
					draw, rectangle split horizontal=false,inner ysep=0pt, line width=0.15em, text width=8em, align=center}}	
			
			\tikzset{rblock/.style = {rectangle split, rectangle split parts=2,
					draw, rectangle split horizontal=false,inner ysep=0pt, line width=0.15em, text width=8em, align=center}}		
			
			\tikzset{b4block/.style = {rectangle split, rectangle split parts=2,
					draw, rectangle split horizontal=false,inner ysep=0pt, line width=0.15em, text width=8em, align=center, rectangle split part fill={olive!30}}}		
			
			\tikzset{b3block/.style = {rectangle split, rectangle split parts=2,
					draw, rectangle split horizontal=false,inner ysep=0pt, line width=0.15em, text width=8em, align=center, rectangle split part fill={orange!60}}}
			
			\tikzset{b2block/.style = {rectangle split, rectangle split parts=1,
					draw, rectangle split horizontal=false,inner ysep=0pt, line width=0.15em, text width=8em, align=center, rectangle split part fill={blue!30}}}	
			
			\node[b4block] (b4) at (3, -3.5) {
				\nodepart{one} {\mystrut $\mathtt{\langle 1, [], \bot \rangle}$ \quad {$1$}}
				\nodepart{two} {\mystrut $\langle 2, [], \bot \rangle$ \quad {$5$}}};
			
			\node[b3block] (b3)  at (-3, -3.5) {
				\nodepart{one} {\mystrut $\mathtt{\langle 1, [], \bot \rangle}$ \quad {$1$}}
				\nodepart{two} {\mystrut $\langle 2, [], \bot \rangle$ \quad {$4$}}};										
			
			\node[b2block] (b2) at (4.5+\y, -1) {
				\nodepart{one} {\mystrut $\mathtt{\langle 1, [30, 40], \bot \rangle}$ \enskip {$3$}}};
			
			
			\node[rblock] (b1)  at (4.5+\y, 1) {
				\nodepart{one} {\mystrut $\mathtt{\langle 1, [10], \bot \rangle}$ \quad {$4$}}
				\nodepart{two} {\mystrut $\mathtt{\langle 2, [10], \bot \rangle}$ \quad {$5$}}};		
			
			
			\node[left, align=left] at (b4.west) {$\qQ_{\mB_4}[1]$};
			\node[left, align=left, xshift=-2em] at (b3.west) {$\qQ_{\mB_3}[1]$};
			\node[left, align=left, yshift=0.75em] at (b3.west) {$\mathtt{30}$};
			\node[left, align=left, yshift=-0.75em] at (b3.west) {$\mathtt{31}$};			
			\node[right, align=right, yshift=0.75em] at (b4.east) {$\mathtt{40}$};
			\node[right, align=right, yshift=-0.75em] at (b4.east) {$\mathtt{41}$};				
			\node[right, align=right] at (b2.east) {$\mathtt{10} \quad \qQ_{\mB_2}[1]$};			
			\node[right, align=right] at (b1.east) {$\qQ_{\mB_1}[()]$};
			
			\end{tikzpicture}
		}
		\caption{Priority queue state (mirroring the decomposition) after preprocessing phase.} \label{fig:preprocessing}
	\end{subfigure}	\hfill 
	\begin{subfigure}[t]{0.99\linewidth}
		\centering
		\scalebox{.99}{\begin{tikzpicture}
			\tikzset{edge/.style = {->,> = latex'},
				vertex/.style={circle, thick, minimum size=7mm}}
			\def\x{0.25}
			\def\y{-1}
			
			\tikzset{pblock/.style = {rectangle split, rectangle split parts=2,
					draw, rectangle split horizontal=false,inner ysep=0pt, line width=0.15em, text width=8em, align=center}}
			
			\tikzset{qblock/.style = {rectangle split, rectangle split parts=1,
					draw, rectangle split horizontal=false,inner ysep=0pt, line width=0.15em, text width=8em, align=center, rectangle split part fill={white}}}	
			
			\tikzset{b4block/.style = {rectangle split, rectangle split parts=2,
					draw, rectangle split horizontal=false,inner ysep=0pt, line width=0.15em, text width=6em, align=center, rectangle split part fill={olive!30}}}		
			
			\tikzset{b3block/.style = {rectangle split, rectangle split parts=2,
					draw, rectangle split horizontal=false,inner ysep=0pt, line width=0.15em, text width=6em, align=center, rectangle split part fill={orange!60}}}
			
			\tikzset{b2block/.style = {rectangle split, rectangle split parts=2,
					draw, rectangle split horizontal=false,inner ysep=0pt, line width=0.15em, text width=8em, align=center, rectangle split part fill={blue!30}}}	
			
			\tikzset{b2pblock/.style = {rectangle split, rectangle split parts=1,
					draw, rectangle split horizontal=false,inner ysep=0pt, line width=0.15em, text width=8em, align=center, rectangle split part fill={blue!30}}}	
			
			\scalebox{.7}{
				
				\node[draw, rectangle, minimum width=12em, minimum height=12.5em, line width=0.1mm, label=above:{\texttt{first popped tuple}}] at (-3.6,-1.1) {};
				
				\node[b2pblock] (b2p) at (-3.5, \y+0.25-1.75) {
					\nodepart{one} {\mystrut $\mathtt{\langle 1, [30, 40], 11 \rangle}$ \enskip {$3$}}};
				
				\node[qblock] (b1p) at (-3.5, 1+\y+0.25) {
					\nodepart[text opacity=0.5, fill opacity=0.5]{one} {\mystrut $\mathtt{\langle 1, [10], \bot \rangle}$ \quad {$4$}}};
				
				\node[left, align=left] at (b2p.west) {$\mathtt{10}   $};
			}

			\node[b4block] (b4) at (6, -3.5+\y) {
				\nodepart{one} {\mystrut $\mathtt{\langle 1, [], 41 \rangle}$ \quad {$1$}}
				\nodepart{two} {\mystrut $\langle 2, [], \bot \rangle$ \quad {$5$}}};
			
			\node[b3block] (b3)  at (-3, -3.5+\y) {
				\nodepart{one} {\mystrut $\mathtt{\langle 1, [], 31 \rangle}$ \quad {$1$}}
				\nodepart{two} {\mystrut $\langle 2, [], \bot \rangle$ \quad {$4$}}};							
			
			
			\node[b2block] (b2) at (1.5, -1+\y) {
				\nodepart{one} {\mystrut $\mathtt{\langle 1, [31, 40], \bot \rangle}$ \enskip {$6$}}
				\nodepart{two} {\mystrut $\mathtt{\langle 1, [30, 41], \bot \rangle}$ \enskip {$7$}}};
			
			%
			
			\node[pblock] (b1)  at (1.5, 1+\y) {
				\nodepart{one} {\mystrut $\mathtt{\langle 2, [10], \bot \rangle}$ \quad {$5$}}
				\nodepart{two} {\mystrut $\mathtt{\langle 1, [11], \bot \rangle}$ \quad {$7$}}};		
			
			
			\node[left, align=right] at (b4.west) {$\qQ_{\mB_4}[1]$};
			\node[right, align=right, yshift=0.75em] at (b4.east) {$\mathtt{40}$};
			\node[right, align=right, yshift=-0.75em] at (b4.east) {$\mathtt{41}$};									
			\node[left, align=left, xshift=-2em] at (b3.west) {$\qQ_{\mB_3}[1]$};
			\node[left, align=left, yshift=0.75em] at (b3.west) {$\mathtt{30}$};
			\node[left, align=left, yshift=-0.75em] at (b3.west) {$\mathtt{31}$};						
			\node[right, align=right] at (b2.east) {$\qquad \qQ_{\mB_2}[1]$};	
			\node[right, align=right, yshift=0.75em] at (b2.east) {$\mathtt{11}$};
			\node[right, align=right, yshift=-0.75em] at (b2.east) {$\mathtt{12}$};						
			\node[right, align=right] at (b1.east) {$\qQ_{\mB_1}[()]$};
			
			\end{tikzpicture}
		}
		\caption{Priority queue state after one iteration of loop in procedure \FuncSty{\texttt{ENUM}()}.} \label{fig:firstenum}
	\end{subfigure}
	\begin{subfigure}[t]{\linewidth}
		\vspace{1em}
		\centering
		\scalebox{.99}{\begin{tikzpicture}
			\tikzset{edge/.style = {->,> = latex'},
				vertex/.style={circle, thick, minimum size=7mm}}
			\def\x{0.25}
			\def\y{-1}
			
			\tikzset{pblock/.style = {rectangle split, rectangle split parts=2,
					draw, rectangle split horizontal=false,inner ysep=0pt, line width=0.15em, text width=6em, align=center}}
			
			\tikzset{qblock/.style = {rectangle split, rectangle split parts=1,
					draw, rectangle split horizontal=false,inner ysep=0pt, line width=0.15em, text width=6em, align=center, rectangle split part fill={lightgray}}}	
			
			\tikzset{b4block/.style = {rectangle split, rectangle split parts=2,
					draw, rectangle split horizontal=false,inner ysep=0pt, line width=0.15em, text width=8em, align=center, rectangle split part fill={olive!30}}}		
			
			\tikzset{b3block/.style = {rectangle split, rectangle split parts=2,
					draw, rectangle split horizontal=false,inner ysep=0pt, line width=0.15em, text width=8em, align=center, rectangle split part fill={orange!60}}}
			
			\tikzset{b2block/.style = {rectangle split, rectangle split parts=2,
					draw, rectangle split horizontal=false,inner ysep=0pt, line width=0.15em, text width=6em, align=center, rectangle split part fill={blue!30}}}	
			
			\tikzset{b2oneblock/.style = {rectangle split, rectangle split parts=1,
					draw, rectangle split horizontal=false,inner ysep=0pt, line width=0.15em, text width=8em, align=center, rectangle split part fill={blue!30}}}	
			
			\tikzset{b2twoblock/.style = {rectangle split, rectangle split parts=1,
					draw, rectangle split horizontal=false,inner ysep=0pt, line width=0.15em, text width=8em, align=center, rectangle split part fill={blue!30}}}
			
			\tikzset{b2threeblock/.style = {rectangle split, rectangle split parts=1,
					draw, rectangle split horizontal=false,inner ysep=0pt, line width=0.15em, text width=8em, align=center, rectangle split part fill={blue!30}}}		
			
			\tikzset{b2fourblock/.style = {rectangle split, rectangle split parts=1,
					draw, rectangle split horizontal=false,inner ysep=0pt, line width=0.15em, text width=8em, align=center, rectangle split part fill={blue!30}}}	
			
			\node[b2oneblock] (b21) at (-4.5, -2) {
				\nodepart{one} {\mystrut $\mathtt{\langle 1, [30, 40], 11\rangle}$ \enskip {$3$}}};
			\node[below, ] at (b21.south) {$\mathtt{10}$};
			
			\node[b2twoblock] (b22) at (4.5, -2) {
				\nodepart{one} {\mystrut $\mathtt{\langle 1, [31, 40], 12\rangle}$ \enskip {$6$}}};
			\node[below, ] at (b22.south) {$\mathtt{11}$};
			
			\node[b2threeblock] (b23) at (-4.5, -3.5) {
				\nodepart{one} {\mystrut $\mathtt{\langle 1, [30,41], 13 \rangle}$ \enskip {$7$}}};
			\node[below, ] at (b23.south) {$\mathtt{12}$};
			
			\node[b2fourblock] (b24) at (4.5, -3.5) {
				\nodepart{one} {\mystrut $\mathtt{\langle 1, [31, 41], \bot \rangle}$ \enskip {$10$}}};
			\node[below, ] at (b24.south) {$\mathtt{13}$};
			
			\node[b4block] (b4) at (4.5, -4.5+\y) {
				\nodepart{one} {\mystrut $\mathtt{\langle 1, [], 41 \rangle}$ \quad {$1$}}
				\nodepart{two} {\mystrut $\mathtt{\langle 2, [], \bot \rangle}$ \quad {$5$}}};
			\node[right, align=right, yshift=0.75em] at (b4.east) {$\mathtt{40}$};
			\node[right, align=right, yshift=-0.75em] at (b4.east) {$\mathtt{41}$};					
			
			\node[b3block] (b3)  at (-4.5, -4.5+\y) {
				\nodepart{one} {\mystrut $\mathtt{\langle 1, [], 31 \rangle}$ \quad {$1$}}
				\nodepart{two} {\mystrut $\mathtt{\langle 2, [], \bot \rangle}$ \quad {$4$}}};		
			\node[left, align=left, yshift=0.75em] at (b3.west) {$\mathtt{30}$};
			\node[left, align=left, yshift=-0.75em] at (b3.west) {$\mathtt{31}$};							
			
			
			\draw[->, color=black] (b21.one east)++(0,0) to[out=0,in=180] (b22.one west);
			\draw[->, color=black] (b22.one east)++(0,0) -- ++(0.3,0) -- (6.50,-2.8) -- (-6.5,-2.8) -- (-6.5,-3.5) --(b23.one west);
			\draw[->, color=black] (b23.one east)++(0,0) to[out=0,in=180] (b24.one west);

			\end{tikzpicture}}
		\caption{The materialized output stored at subtree rooted at $\mB_2$  after enumeration is complete.} \label{fig:materialized}
	\end{subfigure}
	\caption{Preprocessing and enumeration phase for Example~\ref{ex:intro}. Each cell is assigned a memory addressed (written next to the cell). Pointers in cells are populated with the memory address of the cell they are pointing to. Cells are color coded according to the bag (white for root bag, blue for $\mB_{2}$, orange for $\mB_3$ and olive for $\mB_4$.)}
\end{figure}

For the instance shown above and the query decomposition that we have fixed, relation $R_i$ covers bag $\mB_i, i \in [4]$. Each relation has size $N = 2$. Since the relations are already materialized, we only need to perform a full reducer pass, which can be done in linear time. This step removes tuple $(3,1)$ from relation $R_2$ as it does not join with any tuple in $R_1$. 
	
Figure~1(A) shows the state of priority queues after the pre-processing step. For convenience, $\theta$ in each cell $\langle \theta,  [p_1, \dots, p_k], \mnext \rangle$ is shown using the primary key of the tuple and pointers $p_i$ and $\mnext$ are shown using the address of the cell it points to. \hlrone{For example, the tuple $(id=1, \mathbf{w_2}=1, \mathbf{y}=1, \mathbf{z}=1)$ in relation $R_2$ is shown as cell \mytikzcell\ with $\theta = id = 1$ and an empty $\mnext$. The cell in a memory location is followed by the score of the tuple formed by creating the tuple from the pointers in the cell recursively.} For instance, the score of the tuple formed by joining $(\mathbf{y}=1, \mathbf{z}=1) \in R_2$ with $(\mathbf{z}=1, \mathbf{p}=1)$ from $R_3$ and $(\mathbf{z}=1, \mathbf{1}=1)$ in $R_4$ is $1+1+1=3$ (shown as \mytikzcell\ in the figure)\footnote{Note that since our ranking function is sum, we use $v^\star(x) = 0$ for each variable. This allows us to only look at the ``partial" score of tuples that join in a particular subtree.}. The pointer addresses \hlrone{(which have been chosen arbitrarily)} $\mathtt{30}$ and $\mathtt{40}$ refer to the topmost cell in the priority queue for $\mB_3$ and $\mB_4$. Each cell in every priority queue points to the top element of the priority queue of child nodes that it joins with. Note that since both tuples in $R_1$ join with the sole tuple from $R_2$, they point to the same cell.
\end{exa}

\begin{lem} \label{lem:time:space}
	The runtime of Algorithm~\ref{algo:preprocessing} is ${O}(|D|^\mfhw)$. Moreover, at the end of the algorithm, the resulting
	data structure has size $O(|D|^\mfhw)$.
\end{lem}
\begin{proof}
	It is known that the materialization of each bag can be done in time $O(|D|^\fhw)$, and the full reducer pass is linear in the size of the bags~\cite{yannakakis1981algorithms}. For the second step of the preprocessing algorithm, observe that for each valuation in a bag, the algorithm performs only a constant number of operations (the number of children in the tree plus one), where each operation takes a constant time (since insert and top can be done in $O(1)$ time for the priority queue). Hence, the second step needs $O(|D|^\fhw)$ time as well.
	
	Regarding the space requirements, it is easy to see that the data structure uses only constant space for every valuation in each bag, hence the space is bounded by $O(|D|^\fhw)$.
\end{proof}

\introparagraph{Enumeration}
Algorithm~\ref{algo:enumeration} presents the algorithm for the enumeration phase. The heart of the algorithm is the procedure \texttt{TOPDOWN(}$c,t$\texttt{)}.
The key idea of the procedure is that whenever we want to output a new tuple, we can simply obtain it from the top of the priority queue in the root node (node $r$ is the root node of the tree decomposition). Once we do that, we need to update the priority queue by popping the top, and inserting (if necessary) new valuations in the priority queue. This will be recursively propagated in the tree until it reaches the leaf nodes. Observe that as the new candidates are being inserted, the \textsf{next} pointer of cells $c$ at some node of the decomposition are being updated by pointing to the topmost element in the priority queue of its children. This chaining materializes the answers for the particular bag that can be reused.

\begin{exa} \label{ex:enum}
	Figure~\ref{fig:firstenum} shows the state of the data structure after one iteration in \FuncSty{\texttt{ENUM}()}. The first answer returned to the user is the topmost tuple from $\qQ_{\mB_1}[()]$ \hlrone{(shown in the box labeled \textsf{first popped tuple})}.  Cell \myrootcell\ is popped from $\qQ_{\mB_1}[()]$ (after satisfying if condition on line~\ref{line:if} since $\mnext$ is $\bot$). We recursively call \FuncSty{\texttt{TOPDOWN}} for child node $\mB_2$ with cell \mytikzcell\ as the function argument (since that is the cell at memory address $\mathtt{10}$). \hlrone{Recall that \mytikzcell\ was created and pushed into the priority queue $\qQ_{\mB_2}[1]$ during the preprocessing phase.} The \textsf{next} for this cell is also $\bot$ and we pop it from $\qQ_{\mB_2}[1]$. At this point, $\qQ_{\mB_2}[1]$ is empty. The next recursive call is for $\mB_3$ with \mybthreecell\ (cell at memory adress $\mathtt{30}$). The least ranked tuple but larger than \mybthreecell\ in $\qQ_{\mB_3}[1]$ is the cell at address $\mathtt{31}$ . Thus, \textsf{next} for \mybthreecell\ is updated to $\mathtt{31}$ \hlrone{(on line~\ref{chaining})} and cell at memory address $\mathtt{31}$ (which is \mybthreecellnext\ ) is returned, leading to the creation and insertion of \mybtwocell\ cell in $\qQ_{\mB_2}[1]$ \hlrone{on line~\ref{tuple:add}}. Similarly, we get the other cell in $\qQ_{\mB_2}[1]$ after making a recursive call to $\mB_4$. After both the calls are over for node $\mB_2$, the topmost cell at $\qQ_{\mB_2}[1]$ is cell at memory address $\mathtt{11}$ ,which is set as the $\mnext$ \hlrone{(on line~\ref{chaining})} for \mytikzcell\ (changing it into \mytikzcellcopy\ ), terminating one full iteration.

	Let us now look at the second iteration of \FuncSty{\texttt{ENUM}()}. The tuple returned is top element of $Q_{\mB_1}[()]$ which is \mybonecell\ . However, the function \FuncSty{\texttt{TOPDOWN}()} with \mybonecell\ does not recursively go all the way down to leaf nodes. Since \mytikzcellcopy\ already has \textsf{next} populated, we insert \mybonecellcopy\ in $Q_{\mB_1}[()]$ completing the iteration. This demonstrates the benefit of materializing ranked answers at each node in the tree. As the enumeration continues, we are materializing the output of each subtree on-the-fly that can be reused by other tuples in the root bag. 
\end{exa}

New candidates are inserted to the priority queue using the logic on~\autoref{tuple:add} of Algorithm~\ref{algo:enumeration}. Given a bag $s$ with $k$ children $s_1, \dots, s_k$ and a cell $c$, the algorithm increments the pointers $p_1, \dots p_k$ one at a time while keeping the remaining pointers fixed. \eat{Observe that for the tuple $c.\theta$, the candidate generation logic will enumerate the cartesian product  $\times_{i \in [k]} (\sigma_{\key{{s_i}} = c.\theta[\key{{s_i}}]} \tOUT(\subtree{s_i}))$. Here, $\tOUT(\subtree{j})$ is the ranked materialized output of subtree rooted at $\mB_j$ and the selection condition filters only those tuples that agree with $c.\theta$ on the key variables of the children bags.} Indeed as Figure~1(B) shows, initially, only \mytikzcell\ was present in $\qQ_{\mB_2}[1]$ but it generated two cells \mybtwocell\ \hlrone{(observe that \texttt{30} was incremented to \texttt{30} but the second pointer still remains \texttt{40})} and \mybtwocellcopy\ \hlrone{(observe that \texttt{40} was incremented to \texttt{41} but the first pointer still remains \texttt{30})}. When these two cells are popped, they will increment pointers and both of them will generate \mybtwocellcopyfinal. \hlrone{This where the set $H$ helps us by ensuring that duplicates cells are not inserted into the priority queue (via the check on line~\ref{dupli}.} While each cell can generate $k$ new candidates, in the worst-case, each cell can  be generated at most $k$ times and inserted into the priority queue. These duplicates are removed by the check on line~\autoref{dupli}. Since the query size is a constant, we may only need to pop a constant number of times in the worst-case and thus affects the delay guarantee only by a constant factor.

\begin{lem}
	Algorithm~\ref{algo:enumeration} enumerates $Q(D)$ with delay $\delta = {O}(\log |D|)$.
\end{lem}

\begin{proof}
	In order to show the delay guarantee, it suffices to prove that procedure \FuncSty{\textsc{topdown}} takes ${O}(\log |D|)$ time when called from the root node, since getting the top element from the priority queue at the root node takes only $O(1)$ time. 
	
	Indeed, \FuncSty{\textsc{topdown}} traverses the tree decomposition recursively. The key observation is that it visits each node in $\htree$ exactly once. For each node, if $\mnext$ is not $\bot$, the processing takes time $O(1)$. If $\mnext == \bot$, it will perform a constant number of pops -- with cost $O(\log |D|)$ -- and a number of inserts equal to the number of children of the node in the tree $\htree$. Thus, in either case the total time per node is $O(\log |D|)$. Summing up over all nodes in the tree, the total time until the next element is output will be ${O}(\log |D|) \cdot |V(\htree)| = {O}(\log |D|)$.	
\end{proof}

\noindent 
We next bound the space $S_e$ needed by the algorithm during the enumeration phase.

\begin{lem}
	After Algorithm~\ref{algo:enumeration} has enumerated $k$ tuples, the additional space used by the algorithm is  
	$S_e = O(\min \{ k, \vert Q(D) \vert \})$.
\end{lem}

\begin{proof}
	The space requirement of the algorithm during enumeration comes from the size of the priority queues at every bag in the decomposition. Since we have performed a full reducer pass over all bags during the preprocessing phase, and each bag $t$ stores in its priority queues all valuations over $\subtree{t}$, it is straightforward to see that the sum of the sizes of the priorities queues in each bag is bounded by $O(|Q(D)|)$.
	
	To obtain the bound of $O(k)$, we observe that for each tuple that we output, the \FuncSty{\textsc{topdown}}  procedure 
	adds at every node in the decomposition a constant number of new tuples in one of the priority queues in this node (equal to the number of children). Similarly, for the set $H$ that ensures no duplicate cells are added, we also add a constant number of tuples at most. Hence, at most $O(1)$ amount of data will be added in the data structure between two consecutive tuples are output. Thus, if we enumerate $k$ tuples from $Q(D)$, the increase in space will be $k \cdot O(1) = O(k)$.
\end{proof}

\introparagraph{Chaining of cells} Observe that as \FuncSty{\texttt{TOPDOWN}} is called recursively, the $\mnext$ of the cells is continuously being updated. This chaining is critical to achieving good delay guarantees. Intuitively, chaining of cells at a bag allows materialization of the the join result of the subquery rooted at that bag in sorted order. Thus, repeated computation is not being performed and cells at the parent of a bag can re-use the sorted materialization. For example, Figure~1(C) shows the eventual sequence of pointers at node $\mB_2$ which is the ranked materialized output of the subtree rooted at $\mB_2$. The pointers between cells are added to emphasize the chained order. The reader can observe that the score for the cells highlighted in blue are also in increasing order.

Finally, we show that the algorithm correctly enumerates all tuples in $Q(D)$ in sorted order according to the ranking function.

\begin{lem} \label{lem:correct}
	Algorithm~\ref{algo:enumeration} enumerates $Q(D)$ in sorted order according to $\mrf$.
\end{lem}
\begin{proof}
	We will prove our claim by induction on post-order traversal of the decomposition and use the compatibility property of the ranking function with the decomposition at hand. \hlrone{We use $R_s$ to denote the relation corresponding to a node $s$ and $\tOUT(\subtree{s}, u)$ for any non-root node\footnote{For the root node $r$, since $\key{r} = \{\}$, we define $\tOUT(\subtree{r}, \{\}) = \Join_{t \in \subtree{r}} R_t = Q(D)$} to denote the ranked materialized output of $(\Join_{t \in \subtree{s}} (R_t \ltimes u))$, where $u$ is a tuple defined over $\key{s}$ and $\ltimes$ is the standard semijoin operator~\cite{bernstein1981power}. The ranking is done according to the function $\rf_{v^\star_{S}}$ where $S = \nodes \setminus \subtree{s}$. 
 
    We will show that for each node $s$, the algorithm generates $\tOUT(\subtree{s}, u)$ in sorted order according to the function $\rf_{v^\star_{S}}$. Since $\tOUT(\subtree{s}, u)$ is a list, we will frequently use the notation $\tOUT(\subtree{s}, u)[\ell]$ to denote the cell at $\ell^{\text{th}}$ location in the list. First, we prove the following claim.}

    \hlrone{\begin{claim} \label{claim:one}
        For any node $s$ and tuple $u$ defined over $\key{s}$, Algorithm~\ref{algo:enumeration} materializes $\tOUT(\subtree{s}, u)$ in the sorted order according to $\rf_{v^\star_{S}}$ where $S = \nodes \setminus \subtree{s}$. 
    \end{claim}}
	\noindent \introparagraph{Base Case} Let $v^\star$ be the valuation over $\nodes \setminus \mB_s$ according to definition of decomposability. We insert each valuation $\theta$ in the relation $R_{s}$ with score $\rf(v^\star \circ \theta)$ (as shown in line~\ref{pre:insert} of Algorithm~\ref{algo:preprocessing}). \hlrtwo{We now argue that the valuations from the priority queue are popped in the sorted order. Consider two valuations $\theta_1$ and $\theta_2$ that are popped successively. Note that $\theta_1[\key{s}] = \theta_2[\key{s}]$ There are two cases to consider: either $\rf(v^\star \circ \theta_2) > \rf(v^\star \circ \theta_1)$ or $\rf(v^\star \circ \theta_2) = \rf(v^\star \circ \theta_1)$. The first case unambiguously guarantees that $\theta_2 \succeq \theta_1$ since the score for $\theta_1$ is strictly smaller. However, if $\rf(v^\star \circ \theta_2) = \rf(v^\star \circ \theta_1)$, it is not immediately clear whether $\theta_2 \succeq \theta_1$ or $\theta_2 \succeq \theta_1$ because there could be a different valuation $v^\#$ for which $\rf(v^\# \circ \theta_2) \neq \rf(v^\# \circ \theta_1)$. We argue that such a $v^\#$ cannot exist. Indeed,  Definition~\ref{def:decomp} guarantees that if $\rf(v^\star \circ \theta_2) = \rf(v^\star \circ \theta_1)$, then it must also be equal for any other valuation over $\nodes \setminus \mB_s$. In other words, all output tuples $t \in Q(D)$, such that $t[\mB_s] = \theta_1$ or  $t[\mB_s] = \theta_2$ are guaranteed to have the score, and thus, we can safely use the ordering $\theta_2 \succeq \theta_1$ for $\theta_1$ and $\theta_2$.} Since the preprocessing phase already initializes the priority queue, the pop operation will insert the tuple in $\tOUT(\subtree{s}, \theta[\key{s}])$ by populating the $\mnext$ of the cell corresponding to $\theta$ correctly.
	
	\noindent \introparagraph{Inductive Case} Consider some node $s$ in the post-order traversal with children $s_1, \dots s_m$. By the induction hypothesis, the ordering of $\tOUT(\subtree{s_i}, b)$ for each valuation $b$ over $\key{s_i}$ is  generated in sorted order for ranking function $\rf_{v^\star_{S_i}}$ where $S_i = \nodes \setminus  \subtree{s_i}$. Let $\theta$ be a tuple in $R_{s}$ and let $u = \theta[\key{s}]$. Observe that the preprocessing phase creates a cell for $\theta$ whose pointer list $[p_1, \dots p_m]$ is the address of the cell at location $0$ of the materialized list of $\tOUT(\subtree{s_i}, \theta[\key{s_i}])$. We claim that this is the least ranked tuple that can be formed over $\subtree{s}$. Let $c^0_i$ denote the cell at location $0$ for list $\tOUT(\subtree{s_i}, \theta[\key{s_i}])$.  If any pointer $p_i$ points to any other cell (say $d_i$) present at a different location in the list $\tOUT(\subtree{s_i}, \theta[\key{s_i}])$, we can create a smaller ranked tuple by changing $p_i$ to point to the first cell in the list. In other words, since $\out(d_i) \succeq \out(c^0_i)$ (i.e., the first cell is the least ranked, which follows from the correctness of $\tOUT(\subtree{s_i}, \theta[\key{s_i}])$), it holds that

    \begin{align*}
		&\rf(\theta \circ v^\star \circ \out(d_1) \circ \dots\out(d_i) \dots \circ \out(d_m)) \geq \\  &\rf(\theta \circ v^\star \circ \out(d_1) \circ \dots\out(c^0_i) \dots \circ \out(d_m)) \\
	\end{align*}
 
    Here, we use the fact that $\rf$ is $\subtree{s_i} \setminus \key{s_i}$-decomposable conditioned on $\key{s_i}$. Note that no sibling of $s_i$ can have any common variables  with $s_i$ other than the $\key{s}$ which have already been fixed. This proves that the first tuple returned by the pop operation on the priority queue for key $\theta[\key{s}]$ at node $s$ will be correct, which is then added to the list $\tOUT(\subtree{s}, \theta[\key{s}])$.
	
	Next, we proceed to show the correctness for an arbitrary step in the execution. Suppose $c$ is the last cell popped at~\autoref{tuple:pop}. From \autoref{for:1}-\ref{tuple:add}, one may observe that a new candidate is pushed into priority queue for key by incrementing pointers to $\tOUT(\subtree{s_i}, \theta[\key{s_i}])$ one at a time for each child bag $\mB_{s_i}$, while keeping the remainder of the cell content fixed (line~\ref{push:queue}). Let $c.\texttt{MID()}[i]$ point to the cell $\tOUT(\subtree{s_i}, \theta[\key{s_i}])[\ell_i]$. We will use $c_i = \ell_i$ as a shorthand to denote this information. Then, the $m$ candidates generated by the logic will contain pointers that point to the following index location
	
	\begin{align*}
	\mathcal{L} = &\ \ell_1 + 1, \ell_2, \ell_3, \dots, \ell_m, \\
	&\ \ell_1, \ell_2  + 1, \ell_3, \dots, \ell_m, \\
	&\ \ell_1, \ell_2 , \ell_3  + 1, \dots, \ell_m, \\	
	&\ \dots \\
	&\ \ell_1, \ell_2 , \ell_3, \dots, \ell_m+1 \\	
	\end{align*}
	
	\hlrtwo{Suppose that up until now, the algorithm  has generated the ranked output \\$\tOUT(\subtree{s}, \theta[\key{s}])$ in sorted order and the next smallest cell that must be popped on line~\ref{tuple:pop} of Algorithm~\ref{algo:enumeration} is $c^\succ$. Let the pointer at location $i$ in  $c^\succ$.\texttt{MID()} point to index $\ell^\succ_i$ in the list  $\tOUT(\subtree{s_i}, \theta[\key{s_i}])$ (i.e., $c^\succ_i = \ell^\succ_i$). We need to show that $c^\succ$ is a cell with $c^\succ$.\texttt{MID()} as one of the candidates in $\mathcal{L}$ or is already in the priority queue. For the sake of contradiction, suppose there is a cell $c'$ with $c'.\texttt{MID()}[i] = \mathsf{address\_of}( \tOUT(\subtree{s_i}, \theta[\key{s_i})[\ell'_i]), i \in [m]$, that is the next smallest after $c$ but is neither present in $\mL$ and nor present in the priority queue. In other words, we are assuming that $\rf_{v^\star_S}(\out(c')) < \rf_{v^\star_S}(\out(c^\succ))$.\footnote{Note that $\rf_{v^\star_S}(\out(c'))$ cannot be equal to $\rf_{v^\star_S}(\out(c^\succ))$ because otherwise, the order in which the cells are popped from the priority queue can be used to establish the total order (similar to the base case).} We will show that such a scenario will violate the compatibility of the ranking function. There are three possible scenarios regarding the values of $\ell^\succ_i$ and $\ell'_i$.
	\begin{enumerate}
		\item $\ell^\succ_i \leq \ell'_i$ for each $i \in [m]$. This scenario implies that $\rf(v^\star \circ \out(c')) \geq \rf(v^\star \circ \out(c^\succ))$. Indeed, we have that

		\noindent 
		\begin{minipage}{\dimexpr\linewidth-\leftmargin}
		\begin{align*}
		 \rf(v^\star \circ \theta \circ \out(c^\succ_1) \circ  \dots  &\out(c^\succ_m)) \\ &\leq  \rf(v^\star \circ \theta \circ \out(c'_1) \circ \out(c^\succ_2) \circ \dots  \out(c^\succ_m))\\ 
		&\leq  \rf(v^\star \circ \theta \circ \out(c'_1) \circ \out(c'_2) \circ \dots  \out(c^\succ_m)) \\
		& \dots \\
		&\leq  \rf(v^\star \circ \theta \circ \out(c'_1) \circ \out(c'_2) \circ \dots  \out(c'_m)) \\
		\end{align*}
		\end{minipage}

		\noindent 
		Each inequality is a successive application of $(\subtree{s_i} \setminus \key{s_i})$-decomposability since $\out(c'_i) \succeq \out(c^\succ_i)$, which follows from the assumed ordering correctness of $\tOUT(\subtree{s_i}, \theta[\key{s_i}])$. \\Thus, it cannot be the case that $\rf_{v^\star_S}(\out(c')) < \rf_{v^\star_S}(\out(c^\succ))$ without violating the compatibility of the ranking function. \\
		
		\item $\ell^\succ_i \!>\! \ell'_i$ for each $i \!\in \![m]$. This scenario implies that $\rf_{v^\star_S}(\out({c'})\!)\! <\! \rf_{v^\star_S}(\out{(c^\succ)}\!)$ and thus, violates our assumption that all cells ranked smaller than $\out(c^\succ)$ have been generated correctly in sorted order.\\

  \eat{This scenario would mean that $c'$ was generated before $c^\succ$ given our candidate generation logic (line \ref{tuple:pop}-\ref{for:1}), violating our assumption that $c'$ has not been enumerated yet or inserted into the priority queue. Observe that each candidate in $\mL$ dominates the pointer locations of the cell that generated them. Thus, successive applications of the logic on $c'$ will generate $c^\succ$ eventually. Since the preprocessing phase creates a cell with pointer locations $0$, generating $c^\succ$ must happen after $c'$ because the pointers are advanced one at a time and thus cannot go directly from $\ell'_i - 1$ to $\ell'_i + 1$ (and somehow skipping $\ell'_i$) without violating the correctness of priority queue.} 
		
		\item $\ell'$ and $\ell^\succ$ are incomparable. It is easy to see that all candidates in $\mathcal{L}$ dominate\footnote{\hlrone{Given two tuples $t_1$ and $t_2$ defined over the same set of variables $V$, we say that $t_1$ \emph{dominates} $t_2$ if $t_1[v] \geq t_2[v]$ for all $v \in V$.}} pointer locations of $c$ (recall that $c$ is last cell popped from the priority queue) but are incomparable to each other. Also, the only way to generate new candidate tuples is through the logic in line~\ref{for:1}-\ref{tuple:add}. Thus, if $c'$ is not in the priority queue, there are two possibilities. Either there is some cell $c''$ in the priority queue such that $\out{(c'')}$ dominates $\out{(c')}$ and thus, $\rf_{v^\star_S}(\out(c'')) \leq \rf_{v^\star_S}(\out(c'))$. $c''$ will eventually generate $c'$ via a chain of cells that successively dominate each other. As $c$ was popped before $c''$, it follows that $\rf_{v^\star_S}(\out(c)) \leq \rf_{v^\star_S}(\out(c'')) \leq \rf_{v^\star_S}(\out(c'))$, a contradiction to our assumption that $c'$ is the next tuple that must be popped after $c$, which cannot happen until $c''$ is popped. The second possibility is that there is no such $c''$, which will mean that $c^\succ$ and $c'$ are generated in the same for loop on~\autoref{push:queue}. But this would imply that $c'$ is in the priority queue. Thus, both these cases violate one of our assumptions made.
	\end{enumerate}
\noindent 
	Therefore, it cannot be the case that $\rf_{v^\star_S}(\out(c')) < \rf_{v^\star_S}(\out(c^\succ))$ which proves the ordering correctness for node $s$. Recall that for the root node $r$, we have $\key{r} = \{\}$ and thus, we have only a single priority queue. Since Claim~\ref{claim:one} holds for all nodes in the decomposition, we have that  $\tOUT(\subtree{r}, \{\}) = (\Join_{t \in \mB_r} R_t)$ (which is nothing but $Q(D)$) will store the sorted output according to $\rf_{v^\star_S}$ where $S = \nodes \setminus \subtree{r}$. However, since $\nodes = \subtree{r}$, we have that for the root node $\rf_{v^\star_S} = \rf$, as desired. Thus, $Q(D)$ is enumerated in sorted order according to $\rf$.} \qedhere
	
\end{proof}	
\eat{
\begin{proof}
	We will prove our claim by induction on post-order traversal of the decomposition. We will show that the priority queue for each node $s$ gives the output in correct order which in turn populates $\tOUT(\subtree{s})$ in sorted order.
	
	\noindent \introparagraph{Base Case} Correctness for ranked output of $\tOUT(\mB_s)$ for leaf node $s$ is trivial as the leaf node tuples are popped from priority queues in order. Let $\phi^\star$ be the valuation over $\nodes \setminus \mB_s$ according to definition of decomposability. We insert each valuation $\theta$ over node $s$ with score $\rf(\phi^\star \circ \theta )$. Since $\rf$ is compatible with the decomposition, it follows that if $\rf(\phi^\star \circ \theta_2 ) \geq \rf(\phi^\star \circ\theta_1)$ such that $\theta_1[\key{s}] = \theta_1[\key{s}]$, then $\theta_2 \succeq \theta_1$, thus recovering the correct ordering for tuple in $s$.
	
	\noindent Let $s$ be a non-leaf node whose children are leaf nodes $s_1, \dots s_m$. Suppose $\theta$ is a valuation over $\subtree{s}$ popped at~\autoref{tuple:pop}. From \autoref{for:1}-\ref{push:queue}, one may observe that a new candidate is pushed into priority queue for key  by incrementing pointers to materialized output one at a time for each child bag $\mB_{s_i}$, while keeping the remainder of tuple (including $\key{{s_i}}$) fixed (\autoref{push:queue}). Let the notation $\phi^{\succ}$ denotes the smallest tuple in a bag that has rank greater than tuple $\phi$ such that $\phi^{\succ}[\key{\mB}]  = \phi[\key{\mB}]$. Then, $\theta = \theta_1 \circ \theta_2 \circ \dots \circ \theta_m$ (the join of tuples from its $m$ children bags) will {\em generate} the following candidates: 
	\begin{align*}
	   \mathcal{L} = &\ p \circ \theta^{\succ}_1 \circ \theta_2 \circ \dots \circ \theta_m, \\
	&\ p \circ \theta_1 \circ \theta^{\succ}_2 \circ \dots \circ \theta_m, \\
	&\ \dots \\
	&\ p \circ \theta_1 \circ \theta_2 \circ \dots \circ \theta^{\succ}_m
	\end{align*}
	Here $p$ is the projection of $\theta$ over variables in $\mB_s$ but not in any child node (let this set of variables be denoted by $G$). We need to argue that the next smallest valuation after $\theta$ that agrees with $\theta[{\mB_s}]$ is one of the tuples in $\mathcal{L}$ or some candidate already in the priority queue. Suppose there is a tuple $\theta' = \theta'_1 \circ \theta'_2 \circ \dots \circ \theta'_m$ such that $\theta'[G] = \theta[G]$\footnote{If $\theta'[\mB_s] \neq \theta[\mB_s]$, then smallest candidates of $\theta'[\mB_s]$ will be compared with that of $\theta[\mB_s]$ in the parent of $\mB_s$. if they agree on the key but not on other variables in $G$, they will be compared against each other in the priority queue $\qQ_s$, which will correctly put them in heap order given their scores.}. For the sake of contradiction, suppose $\theta'$ has strictly smaller score than any tuple in $\mathcal{L}$ or $\theta$, but has not been enumerated yet. In other words, $\rf(\phi \circ \theta') < \rf(\phi \circ \theta)$ for any valuation $\phi$ over $\nodes \setminus \subtree{s}$. Clearly, since $\theta'$ has smaller rank than $\theta$ but has not been enumerated, it follows that $\theta'$ has not been inserted into the priority queue. We will show that such a scenario will violate compatibility of ranking function. Recall that $\theta_1$ is said to dominate $\theta_2$ whenever $\theta_1(x) \succeq \theta_2(x)$ for every variable $x \in S$ implies $\rf(\theta_1) \geq \rf(\theta_2)$. There are three possible scenarios regarding the tuples $\theta$ and $\theta'$:
	
	\begin{enumerate}
		
		\item 	$\theta$ dominates $\theta'$ over each variable in $S_i = \subtree{{s_i}} \setminus \key{{s_i}},$ $i\in [m]$. Note that this scenario would mean that $\theta'$ was generated before $\theta$ given our candidate generation logic (line \ref{tuple:pop}-\ref{for:1}), violating our assumption that $\theta'$ has not been generated yet and inserted into the priority queue. Observe that each candidate in $\mL$ dominates the tuple that generated them. Thus, successive applications of the logic on $\theta'$ will generate $\theta$.
		\\
		
		\item $\theta'$ dominates $\theta$ over each variable in $S_i$. This scenario implies that $\rf(\phi \circ \theta) \leq \rf(\phi \circ \theta')$. We have that,

		\begin{align*}
		 \qquad \rf(\phi \circ p \circ \theta_1 \circ  \dots  \theta_m) &\leq  \rf(\phi \circ p \circ \theta'_1 \circ \theta_2 \ \dots  \theta_m) \\
		&\leq  \rf(\phi \circ p \circ \theta'_1 \circ \theta'_2 \dots \theta_m) \\
		& \dots \\
		&\leq  \rf(\phi \circ p \circ \theta'_1 \circ \theta'_2 \dots \theta'_m) \\
		\end{align*}
		Each inequality is a successive application of $(\subtree{s_i} \setminus \key{s_i})$-decomposability since $\theta'_i \succeq \theta_i$ by domination assumption of $\theta'$ over $\theta$.  \\
		
		\item $\theta'$ and $\theta$ are incomparable. It is easy to see that all candidates in $\mathcal{L}$ dominate $\theta$ but are incomparable to each other. Also, the only way to generate new candidate tuples is~\autoref{for:1}-\ref{push:queue}. Thus, if $\theta'$ is not in the priority queue, there are two possibilities. Either there is some tuple $\theta''$ in the priority queue that is dominated by $\theta'$ and thus, $\rf(\theta'') \leq \rf(\theta')$. $\theta''$ will eventually generate $\theta'$ via a chain of tuples that successively dominate each other. As $\theta$ was popped before $\theta''$, it follows that $\rf(\theta) \leq \rf(\theta'') \leq \rf(\theta')$, a contradiction to our assumption that $\theta'$ has a smaller rank than $\theta$. The second possibility is that there is no such $\theta''$, which will mean that $\theta$ and $\theta'$ are generated in the same for loop~\autoref{for:1}. But this would again mean that $\theta'$ is in the priority queue. Thus, both these cases violate our assumption that $\rf(\theta') < \rf(\theta)$.
		
	\end{enumerate}	
	
	Therefore, it cannot be the case that $\rf(\phi \circ \theta') < \rf(\phi \circ \theta^\succ)$ which proves the ordering correctness for node $s$. Since the output $\tOUT(\subtree{s})$ is populated using this ordering form priority queue, it is also materialized (chaining of cells at~\autoref{chaining}) in ranked order.
	
	\noindent \introparagraph{Inductive Case} Consider some node $s$ in post-order traversal with children $s_1, \dots s_m$. By induction hypothesis, the ordering of $\tOUT(\mB_{s_i})$ and correctness of $\qQ_{s_i}$ is guaranteed. Applying the same argument as in the base case, it is straightforward to show the correctness for bag $s$. This completes the proof.
	
	It is easy to see that the algorithm indeed enumerates all tuples in $Q(D)$ since the full reducer pass removes all dangling tuples. 
	
\end{proof}	
}

	\section{Extensions} \label{sec:extension}

In this section, we describe two extensions of \autoref{thm:main} and how it can be used to further improve the main result. 

\subsection{Ranked Enumeration of UCQs}

We begin by discussing how ranked enumeration of full UCQs can be done. The first observation is that given a full UCQ $\varphi = \varphi_1 \cup \dots \varphi_\ell$, if the ranked enumeration of each $\varphi_i$ can be performed efficiently, then we can perform ranked enumeration for the union of query results. This can be achieved by applying \autoref{thm:main} to each $\varphi_i$ and introducing another priority queue that compares the score of the answer tuples of each $\varphi_i$, pops the smallest result, and fetches the next smallest tuple from the data structure of $\varphi_i$ accordingly. Although each $\varphi_i(D)$ does not contain duplicates, it may be the case that the same tuple is generated by multiple $\varphi_i$. Thus, we need to introduce a mechanism to ensure that all tuples with the same weight are enumerated in a specific order. Fortunately, this is easy to accomplish by modifying Algorithm~\ref{algo:enumeration} to enumerate all tuples with the same score in lexicographic increasing order. The choice of lexicographic ordering as a tie-breaking criteria is not the only valid choice. As long as the ties are broken consistently, other ranking functions can also be used. This ensures that tuples from each $\varphi_i$ also arrive in the same order. Since each $\varphi_i$ is enumerable in ranked order with delay $O(\log|D|)$ and the overhead of the priority queue is $O(\ell)$ (priority queue contains at most one tuple from each $\varphi_i$), the total delay guarantee is bounded by $O(\ell \cdot \log|D|) = O(\log|D|)$ as the query size is a constant.  The space usage is determined by the largest fractional hypertree-width across all decompositions of subqueries in $\varphi$. This immediately leads to the following result.

\begin{thm} \label{thm:ucq}
	Let $\varphi = \varphi_1 \cup \dots \varphi_\ell$ be a full UCQ. Let \fhw\ denote the fractional hypertree-width of all decompositions across all CQs $\varphi_i$, and $\rf$ be a ranking function that is compatible with the decomposition of each $\varphi_i$. Then, for any input database $D$, we can pre-process $D$ in time and space,
	$$
	T_p = {O}(|D|^{\fhw}) \quad \quad S_p = {O}(|D|^{\fhw})
	$$
	such that for any $k$, we can enumerate the top-$k$ tuples of $\varphi(D)$ with
	$$
	\text{delay } \delta = {O}(\log |D|)  \quad \quad
	\text{space } S_e = O(\min \{k, \vert \varphi(D) \vert\})
	$$
	
\end{thm}
\vspace{0.5\baselineskip}
\noindent 
Algorithm~\ref{algo:subw} shows the enumeration algorithm. It outputs one output tuple $t$ in every iteration and line~\ref{line:one}-\ref{line:two} pop out all duplicates of $t$ in the queue. Recall that since $Q = Q_1 \cup \dots Q_\ell$, there can be at most $\ell$ duplicates for some constant $\ell$. \FuncSty{$\texttt{ENUM}(Q_i)$} is the invocation of \FuncSty{$\texttt{ENUM}()$} procedure from Algorithm~\ref{algo:enumeration} on query $Q_i$.

\begin{algorithm}
	\SetCommentSty{textsf}
	\DontPrintSemicolon 
	\SetKwFunction{fullreducer}{\textsc{Materialize}}
	\SetKwFunction{initializepq}{\textsc{InitializePQ}}
	\SetKwFunction{fillout}{\textsc{topdown}}
	\SetKwFunction{ins}{\textsc{insert}}
	\SetKwFunction{app}{\textsc{insert}}
	\SetKwFunction{top}{\textsc{top()}}
	\SetKwFunction{pop}{\textsc{pop()}}
	\SetKwFunction{push}{\textsc{push}}
	
	\SetKwFunction{enum}{\textsc{enum}}	
	\SetKwFunction{preprocess}{\textsc{preprocess}}	
	\SetKwFunction{eoe}{\textsc{eoe}}	
	\SetKwFunction{queue}{\textsc{Queue}}	
	\SetKwData{pointer}{\textsf{pointer}}
	\SetKwData{temp}{\textsf{temp}}
	\SetKwData{retval}{\textsf{retval}}
	\SetKwData{popvalue}{\textsf{popvalue}}
	\SetKwInOut{Input}{\textsc{input}}\SetKwInOut{Output}{\textsc{output}}
	\SetKwProg{myproc}{\textsc{procedure}}{}{}
    \hlrone{\Input{Full UCQ $Q = Q_1 \cup \dots \cup Q_\ell$, Tree decomposition $(\htree_i, \mB^i_{t \in V(\htree_i)})$ for each CQ $Q_i$, Database $D$, Ranking function $\rf$}}
     \hlrone{\Output{Enumerate $Q(D)$ in sorted order according to $\rf$}}
	\myproc{\preprocess{$Q, D, \mrf, (\htree_i, \mB^i_{t \in V(\htree_i)})$}}{
		Apply Algorithm~\ref{algo:preprocessing} to all $Q_i$ \;
		$\queue \leftarrow \emptyset$ \;
		\For{$i \in \{1, \dots, \ell\}$}{
			$\queue.\push(\enum(Q_i))$ \tcc*{Initialize $\queue$ with smallest candidate for each $Q_i$}
		}
	}
	
	\myproc{\enum{$Q, D, \mrf, (\htree_i, \mB^i_{t \in V(\htree_i)})$} \label{func:enum}}{
		\While{$\queue$ is not empty}{
			$t \gets \queue.\pop$ \;
			\textbf{output} $t$ \tcc*{Suppose  $t$ came from subquery $Q_i$}
			$\queue.\push(\enum(Q_i))$ \tcc*{Push the next candidate for $Q_i$}
			\While{$\queue.\top == t$}{
				$\queue.\pop$ \label{line:one} \tcc*{drain the queue of duplicate $t$}
				\tcc{Suppose duplicate $t$ came from calling $\enum(Q_j)$ on subquery $Q_j$}
				$\queue.\push(\enum(Q_j))$ \label{line:two} \tcc*{Push the next candidate for $Q_j$}
			}
		}
	}
	\caption{Preprocessing and Enumeration Phase} \label{algo:subw}
\end{algorithm}

The comparison function for priority queues in Algorithm~\ref{algo:enumeration} for each subquery $Q_i$ of $Q$ is modified in the following way. Consider two tuples $t_1$ and $t_2$ with schema $(x_1, x_2, \dots, x_n)$ and scores $\rank{t_1}$ and $\rank{t_2}$ respectively.

\begin{algorithm}
	\SetCommentSty{textsf}
	\DontPrintSemicolon 
	\SetKwFunction{fullreducer}{\textsc{Materialize}}
	\SetKwFunction{initializepq}{\textsc{InitializePQ}}
	\SetKwFunction{fillout}{\textsc{topdown}}
	\SetKwFunction{ins}{\textsc{insert}}
	\SetKwFunction{app}{\textsc{insert}}
	\SetKwFunction{top}{\textsc{top()}}
	\SetKwFunction{pop}{\textsc{pop()}}
	\SetKwFunction{push}{\textsc{pop}}
	
	\SetKwFunction{compare}{\textsc{compare}}	
	\SetKwFunction{eoe}{\textsc{eoe}}	
	\SetKwFunction{queue}{\textsc{Queue}}	
	\SetKwData{pointer}{\textsf{pointer}}
	\SetKwData{temp}{\textsf{temp}}
	\SetKwData{retval}{\textsf{retval}}
	\SetKwData{popvalue}{\textsf{popvalue}}
	\SetKwInOut{Input}{\textsc{input}}\SetKwInOut{Output}{\textsc{output}}
	\SetKwProg{myproc}{\textsc{procedure}}{}{}
    \hlrone{\Input{Tuples $t_1$ and $t_2$, Ranking function $\rf$}}
    \hlrone{\Output{Returns the smaller ranked tuple of $t_1$ and $t_2$; break ties in lexicographic ordering}}
	\myproc{\compare$(t_1, t_2)$}{
		\If{$\mrf(t_1) < \mrf(t_2)$}{
			\texttt{return} $t_1$ \;
		}
		\If{$\mrf(t_2) < \mrf(t_1)$}{
			\texttt{return} $t_2$ \;
		}
		\ForEach{$i \in \{n, n-1, \dots, 1\}$}{
			\If{$\pi_{x_i}(t_1) < \pi_{x_i}(t_2)$}{\texttt{return} $t_1$}
			\If{$\pi_{x_i}(t_2) < \pi_{x_i}(t_1)$}{\texttt{return} $t_2$}
		}
	}
	\caption{Comparison function for priority queues} \label{comparison}
\end{algorithm}

Comparison function in Algorithm~\ref{comparison} compares $t_1$ and $t_2$ based on the ranking function and tie breaks by using the lexicographic ordering of the two tuples. This ensures that all tuples with the same score arrive in a fixed from \FuncSty{$\texttt{ENUM}(Q_i)$} procedure of each subquery $Q_i$.

\subsection{Improving The Main Result}

Although \autoref{thm:ucq} is a straightforward extension of \autoref{thm:main}, it is powerful enough to improve the pre-processing time and space of \autoref{thm:main} by using \autoref{thm:ucq} in conjunction with \emph{data-dependent} tree decompositions. It is well known that the query result for any CQ can be answered in time ${O}(|D|^\fhw + |Q(D)|)$ time and this is asymptotically tight~\cite{AGM}. However, there exists another notion of width known as the \emph{submodular width} (denoted \subw)~\cite{marx2013tractable}. It is also known that for any CQ, it holds that $\subw \leq \fhw$. Recent work by Abo Khamis et al.~\cite{abo2017shannon} presented an elegant algorithm called \panda\ that constructs multiple decompositions by partitioning the input database to minimize the intermediate join size result. \panda\ computes the output of any full CQ in time ${O}(|D|^\subw \cdot \log|D| + |\tOUT|)$. In other words, \panda\ takes a CQ query $Q$ and a database $D$ as input and produces multiple tree decompositions in time ${O}(|D|^\subw\cdot\log|D|)$ such that each answer tuple is generated by at least one decomposition. The number of decompositions depends only on size of the query and not on $D$. Thus, when the query size is a constant, the number of decompositions constructed is also a constant. We can now apply \autoref{thm:ucq} by setting $\varphi_i$ as the tree decompositions produced by \panda\ to get the following result.

\begin{thm} \label{thm:main:subw}
	Let $\varphi$ be a natural join query with hypergraph $\mH = (\nodes, \edges)$, submodular width \subw, and $\mrf$ be a ranking function that is compatible with each tree decomposition of $\varphi$.
	Then, for any input database $D$, we can pre-process $D$ in time and space,
	$$
	T_p = {O}(|D|^{\subw} \cdot \log |D|) \quad \quad S_p = {O}(|D|^{\subw})
	$$
	such that for any $k$, we can enumerate the top-$k$ tuples of $\varphi(D)$ with
	$$
	\text{delay } \delta = {O}(\log |D|)  \quad \quad
	\text{space } S_e = O(\min \{k, \vert \varphi(D) \vert\})
	$$
\end{thm}
	

\section{Lower Bounds}
\label{sec:lowerbound}

In this section, we provide evidence for the near optimality of our results. 

\subsection{The Choice of Ranking Function}
We first consider the impact of the ranking function on the performance of ranked enumeration. 
We start with a simple observation that deals with the case where $\rf$ can be accessed only through a blackbox that, given a tuple/valuation, returns its score: we call this 
a {\em blackbox} \footnote{Blackbox implies that the score $\rf(\theta)$ is revealed only upon querying the function.} ranking function. Note that all of our algorithms work under the blackbox assumption.

\begin{prop}
Let $Q$ be a natural join query, and $\mrf$ be an arbitrary blackbox ranking function. Then, any enumeration algorithm
on a database $D$ needs $\Omega(\vert Q(D) \vert)$ calls to $\mrf$ in order to output the smallest tuple.
\end{prop}
\hlrone{\begin{proof}
    Suppose that some algorithm returns $t \in Q(D)$ as the least ranked tuple without examining the rank of tuple $t^\star \in Q(D)$. Then, the ranking function can assign a rank to $t^\star$ such that $\rf(t^\star) < \rf(t)$. Therefore, any algorithm must examine the rank of each tuple in the output of the query before returning the smallest tuple.
\end{proof}}
\noindent 
\hlrone{The above proposition shows that without any additional restrictions on the ranking function}, 
the simple result in Proposition~\ref{prop:simple:enum} that materializes and sorts the output is essentially optimal.
Thus, it is necessary to exploit properties of the ranking function in order to construct better algorithms. Unfortunately, even for certain natural restrictions of ranking functions, it is not possible to do much better than the
$|D|^{\rho^*}$ bound for certain queries. 

One such a natural restriction is that of coordinate-decomposable functions, 
where we can show the following lower bound result:

\begin{lem} \label{lem:lowerbound}
Consider the query $Q (x_1,y_1, x_2, y_2) = \ R(x_1,y_1), S(x_2, y_2)$
and suppose $\mrf$ is a blackbox ranking function that \hlrtwo{is also known to be coordinate-decomposable}. Then, there exists an instance of size $N$
such that the time required to find the smallest  tuple is $\Omega(N^2)$.
\end{lem}
\begin{proof}
	We construct an instance $D$ as follows. For every variable we use the domain
	$\{a_1, \dots, a_N\}$, which we equip with the order $a_1 < a_2 < \dots < a_N$.
	Then, every tuple in $R$ and $S$ is of the form $(a_i, a_{N-i+1})$ for $i = 1, \dots, N$.
	Similarly, every tuple in $S$ is of the form $(a_i, a_{N-i+1})$.
	
	To construct a family of coordinate-decomposable ranking functions, we consider all ranking functions
	that are monotone w.r.t. the order of the domain $\{a_1, \dots, a_N\}$ for every variable.
	
	
	We will show that any two tuples in $Q(D)$ are incomparable, in the sense that neither tuple dominates the other in all
	variables. Indeed, consider two distinct tuples $t_1 = ( a_i, a_{N-i+1}, a_j, a_{N-j+1} )$, and $t_2 = ( a_k, a_{N-k+1}, a_\ell, a_{N-\ell+1} )$.
	For the sake of contradiction, suppose $t_1$ dominates $t_2$. Then, we must have $i \geq k$ and $N-i+1 \geq N-k+1$, giving $i=k$. Similarly, $j = \ell$.  But this contradicts our assumption that $t_1 \neq t_2$. 
	
	Therefore, a ranking function from our family can assign an arbitrary score to the $N^2$ tuples without violating the coordinate decomposability. \hlrone{Indeed, coordinate decomposability tells us that for any two $\theta_1, \theta_2 \in Q(D)$ that agree on the three variables $\{x_1,y_1, x_2, y_2\} \setminus z$ for any $z \in \{x_1,y_1, x_2, y_2\}$, if $\theta_1(z) \succeq \theta_2(z)$, then $\rf(\theta_1) \geq \rf(\theta_2)$. In other words, the ranking function imposes a constraint on the score only if $\theta_1$ dominates $\theta_2$ or vice-versa.}
    Thus, for any non-dominating tuple pair, the ranking function is free to assign any value as the score. Applying \autoref{lem:lowerbound} gives us the desired lower bound.
\end{proof}

\autoref{lem:lowerbound} shows that for coordinate-decomposable functions, there exist queries where  obtaining constant (or almost constant) delay requires the algorithm to spend superlinear time during the preprocessing step. 
Given this result, the immediate question is to see whether we can extend the lower bound to other CQs. \hlrtwo{We first show a simple but powerful result for coordinate-decomposable functions. Before we present the result, we need to formally define the notion of path and diameter in a hypergraph.

\begin{defi}
	Given a connected hypergraph $\mH = (\nodes, \edges)$, a path $P$ in $\mH$ from vertex $x_1$ to $x_{s+1}$ is a vertex-edge alternate set $x_1E_1x_2E_2 \dots x_s E_s x_{s+1}$ such that $\{x_i, x_{i+1}\} \subseteq E_i (i \in [s])$ and $x_i \neq x_{j}, E_i \neq E_j$ for $i \neq j$. Here, $s$ is the length of the path $P$. The distance between any two vertices $u$ and $v$, denoted $d(u,v)$, is the length of the shortest path connecting $u$ and $v$. The {\em diameter} of a hypergraph, $\dia(\mH)$, is the maximum distance between all pairs of vertices.
\end{defi}

\begin{lem} \label{lem:vertex:binary}
Consider a full acyclic connected query $Q$ over binary relations and $\mrf$ a blackbox ranking function that is also known to be coordinate-decomposable. Then, there  exists an algorithm that enumerates the result of $Q$ in ranked order with $O(\log |D|)$ delay guarantee and $T_p = O(|D|)$ preprocessing time if and only if $\dia(Q) \leq 3$.
\end{lem}
\begin{proof}

First, note that if $\dia(Q) \geq 4$, then there exists a path of the form $x_1 R y_1 T z_1 U x_2 S y_2$. We can embed the hard instance from \autoref{lem:lowerbound} in the following way. We use the same relations as defined in \autoref{lem:lowerbound} to create relations $R$ and $S$. Let the domain of $z_1$ be $z^\star$. We define $T(y_1,z_1) = \{(a_i, z^\star) \;|\; i \in [N]\}$ and $U(z_1,x_1) = \{(z^\star, a_i) \;|\; i \in [N]\}$. For all other relations (let us use $W$ to denote such a relation) in the query other than $R,S,T$ and $U$, we create an instance in the following way. Fix the domain of all variables other than $x_1,y_1,x_2,y_2$ as $a^\star$. Then, $W(p,q) = \{ \domain(p) \times \domain(q) \}$. It is easy to see that size of the relation is at most $N$ since only one of the variables can be $x_1,y_1,x_2,y_2$ (otherwise the query becomes cyclic) and the output of the query will be non-empty. Using the same argument as before, we get $\Omega(N^2) = \Omega(|D|^2)$ incomparable tuples, giving us a lower bound of $\Omega(|D|^2)$ to find the least ranked tuple. 

If $\dia(Q) \leq 3$, we will show that there exists a join tree of depth one. Indeed, if there exists no join tree of depth one, then there exists a root to leaf path (say root node $r$, its child $s$, and child of $s$ as $t$) of length two. The root node must also have at least two children because otherwise, one could make $s$ as the root with $r$ and $t$ as its children. Let the other child of the root node be $u$. We also note that each leaf node bag contains exactly one variable in common with the parent and one variable that is unique to the bag of the leaf node (i.e., it does not appear in any other bag). Thus, the distance from the unique variable in $\mB_t$ to the unique variable in $\mB_u$ requires traversing all of the three intermediate nodes, which leads to a shortest path of length four, a contradiction. Thus, there must exists a join tree of depth one. Next, we show the compatibility of the ranking function with the join tree.  The root bag is $\mbroot$-decomposable by definition since $\key{r} = \{\}$. Let $u_i$ be the unique variable in bag $\mB_i$ for node $s_i$. Then, $\htree$ is $u_i$-decomposable conditioned on $\key{s_i}$. Indeed, since the ranking functions is $u_i$-decomposable, we can apply Proposition~\ref{lem:conditioned} by fixing $T = \{u_i\}$ and $S = \key{s_i} \subseteq \nodes \setminus \{u_i\}$ to obtain the desired result. Thus, \autoref{thm:main} is applicable.
\end{proof}
\noindent 
Our next result characterizes a class of queries which admit efficient ranked enumeration for edge-decomposable ranking functions: these are functions that
 are $S$-decomposable for any $S$ that is a hyperedge in the query hypergraph. 
\eat{Before we present the result, we need to formally define the notion of path and diameter in a hypergraph.

\begin{defi}
	Given a connected hypergraph $\mH = (\nodes, \edges)$, a path $P$ in $\mH$ from vertex $x_1$ to $x_{s+1}$ is a vertex-edge alternate set $x_1E_1x_2E_2 \dots x_s E_s x_{s+1}$ such that $\{x_i, x_{i+1}\} \subseteq E_i (i \in [s])$ and $x_i \neq x_{j}, E_i \neq E_j$ for $i \neq j$. Here, $s$ is the length of the path $P$. The distance between any two vertices $u$ and $v$, denoted $d(u,v)$, is the length of the shortest path connecting $u$ and $v$. The {\em diameter} of a hypergraph, $\dia(\mH)$, is the maximum distance between all pairs of vertices.
\end{defi}}

\begin{lem} \label{thm:edge:dichotomy}
Consider a full acyclic query $Q$ and a blackbox ranking function $\rf$ that is also known to be edge-decomposable. Then, if $Q$ admits a join tree of depth one, then there exists an algorithm that enumerates the result of $Q$ in ranked order with $O(\log |D|)$ delay and $T_p = O(|D|)$ preprocessing time.
\end{lem}
\begin{proof}
	\eat{To prove the hardness result, suppose that $\dia(Q) \geq 4$.
	We construct a database instance $D$ and a family of edge-decomposable ranking functions such that any algorithm must make $\Omega(|D|^2)$ calls to the ranking function to output the first tuple in $Q(D)$.
	Indeed, since $\dia(Q) \geq 4$, there exists a path $x_1 R_1 y_1 S_1 z S_2 y_2 R_2 x_2$ in the hypergraph of $Q$.
	Since this path cannot be made shorter, this implies that $x_1 \notin S_1, S_2, R_2$; $y_1 \notin S_2, R_2$; $z \notin R_1, R_2$; $y_2 \notin R_1, S_1$ and $x_2 \notin R_1, S_1, S_2$.
	
	Figure~\ref{fig:lowerbound} shows how the construction for the database instance $D$; 
	all variables not depicted take the same constant value. 
	For the family of ranking functions, we consider all functions that are monotone with respect to the order of the tuples as depicted in Figure~\ref{fig:lowerbound}. Using the same argument as in \autoref{lem:lowerbound}, it is easy to see that any algorithm must examine the rank of $\Omega(n^2)$ tuples in order to find the smallest one.}

    Consider a join tree of depth one for $Q$. We will show that any such decomposition is compatible with an edge-decomposable function. First, note that the for the root node $r$, any ranking function is $\subtree{r}$-decomposable since $\key{r} = \{\}$. Consider a child node $s$ of the root. Since $\rf$ is edge-decomposable, it implies that for node $s$, the decomposition is $\mB_s$-decomposable. Recall that if a ranking function is $(S \cup T)$-decomposable, then it is also $T$-decomposable conditioned on $S$. Since $\mB_s = (\mB_s \setminus \key{s}) \cup \key{s}$ and $\subtree{s} = \mB_s$ as node $s$ is a leaf, we get that for any leaf node $s$, the ranking function is $(\subtree{s} \setminus \key{s})$-decomposable conditioned on $\key{s}$. Since all nodes in the decomposition are either leaf or the root, we get the compatability of the ranking function with the decomposition at hand.
\end{proof}
\begin{figure}[!t]
	\centering
	\scalebox{1}{
		\begin{tikzpicture}
		\tikzset{vertex/.style={circle, fill=black!25, minimum size=8pt}}
		
		\node[color=red] at (-4.5, 1) {$R_1(x_1,y_1)$};
		\node[color=blue] at (-2, 1) {$R_2(y_1,z)$};
		\node[color=orange] at (0.5, 1) {$R_3(y_2,z)$};
		\node[color=violet] at (3, 1) {$R_4(x_2,y_2)$};
		
		\node[vertex] (X-1) at (-5,0) {$a_1$};
		\node[vertex] (Y-1) at (-3,0) {$b_1$};
		\node[vertex] (X-2) at (-5,-1) {$a_2$};
		\node[vertex] (Y-2) at (-3,-1) {$b_2$};			
		\node[draw=none] at (-4,-1.5) {$\boldsymbol{\vdots}$};
		\node[vertex] (X-n) at (-5,-2.5) {$a_n$};
		\node[vertex] (Y-n) at (-3,-2.5) {$b_n$};		
		
		\node[vertex] (Z) at (-1,-1.25) {$c$};		
		
		\node[vertex] (W-1) at (1,0) {$d_1$};
		\node[vertex] (U-1) at (3,0) {$e_1$};
		\node[vertex] (W-2) at (1,-1) {$d_2$};
		\node[vertex] (U-2) at (3,-1) {$e_2$};			
		\node[draw=none] at (2,-1.5) {$\boldsymbol{\vdots}$};
		\node[vertex] (W-n) at (1,-2.5) {$d_n$};
		\node[vertex] (U-n) at (3,-2.5) {$e_n$};		
		
		\draw[color=red, line width=0.2mm] (X-1) -- node [above, color=black, xshift=0em, yshift=0em] {$1$} (Y-1);
		\draw[color=red, line width=0.2mm] (X-2) -- node [above, color=black, xshift=0em, yshift=0em] {$2$} (Y-2);
		\draw[color=red, line width=0.2mm] (X-n) -- node [above, color=black, xshift=0em, yshift=0em] {$n$} (Y-n);
		\draw[color=violet, line width=0.2mm] (W-1) -- node [above, color=black,  xshift=0em, yshift=0em] {$1$} (U-1);
		\draw[color=violet, line width=0.2mm] (W-2) -- node [above, color=black,  xshift=0em, yshift=0em] {$2$} (U-2);
		\draw[color=violet, line width=0.2mm] (W-n) -- node [above, color=black,  xshift=0em, yshift=0em] {$n$} (U-n);		
		\draw[color=blue, line width=0.2mm] (Y-1) -- node [above, color=black,  xshift=0em, yshift=0em] {$n$}  (Z);
		\draw[color=blue, line width=0.2mm] (Y-2) -- node [above, color=black,  xshift=0em, yshift=0em] {$n-1$}  (Z);
		\draw[color=blue, line width=0.2mm] (Y-n) -- node [above, color=black,  xshift=0em, yshift=0em] {$1$}  (Z);
		\draw[color=orange, line width=0.2mm] (W-1) -- node [above, color=black,  xshift=0em, yshift=0em] {$n$}  (Z);
		\draw[color=orange, line width=0.2mm] (W-2) -- node [above, color=black,  xshift=0em, yshift=0em] {$n-1$}  (Z);
		\draw[color=orange, line width=0.2mm] (W-n) -- node [above, color=black,  xshift=0em, yshift=0em] {$1$}  (Z);
		\end{tikzpicture}
	}
	\caption{Database instance $D$ for the 4-path query. Each edge is color coded by the relation it belongs to. Values over the edges denote the weight assigned to each tuple.}	\label{fig:lowerbound}
\end{figure}

\noindent 
As an example, $Q(x,y,z,w) = R(x,y),S(y,z),T(z,w)$ has a decomposition of depth one where $\{y,z\}$ is the root and $\{x,y\}$ and $\{z,w\}$ are the leaves, and thus we can enumerate the result with linear preprocessing time and logarithmic delay for any edge-decomposable ranking function. 

On the other hand, for the 4-path query $Q(x,y,z,w,t) = R(x,y),S(y,z),T(z,w),U(w,t)$, it is not possible to achieve this. Figure~\ref{fig:lowerbound} shows a database instance with $n$ tuples for the 4-path query. For the family of ranking functions, we consider all functions that are monotone with respect to the order of the tuples as depicted in the figure. Using the same argument as in \autoref{lem:lowerbound}, it is easy to see that any algorithm must examine the rank of $\Omega(n^2)$ tuples in order to find the smallest one. Our last result of this section extends the idea to show a dichotomy for queries over binary relations with edge-decomposable functions.

\begin{lem} \label{thm:edge:dichotomy:binary}
Consider a full acyclic query $Q$ over binary relations and a blackbox ranking function $\rf$ that is also known to be edge-decomposable. Then, there exists an algorithm that enumerates the result of $Q$ in ranked order with $O(\log |D|)$ delay and $T_p = O(|D|)$ preprocessing time if and only if $Q$ admits a join tree of depth one.
\end{lem}
\begin{proof}
    For the one direction, \autoref{thm:edge:dichotomy} already shows the desired result (for all full acyclic CQs, and not just for binary relations) if there exists a join tree of depth one.

    For the other direction, suppose that $Q$ does not have a join tree of depth one. Then, we claim that there is a connected component in the hypergraph of $Q$ with diameter at least four or there exist at least two connected components, each with diameter two or more. Indeed, if the connected components all have diameter one (i.e. each component only has one relation), we can pick any relation as the root and all other relations can become the leaf. Similarly, if there exists a component $C$ with diameter two or three three, and all other components have diameter one, then the isolated relations can directly be made as the children of the root node in the join tree of $C$ (which is guaranteed to be of depth one).
    
    Consider the connected component $Q'$ with diameter at least four. Then, there must exist a path of the form $x_1 R_1 y_1 R_2 z R_3 y_2 R_4 x_2$ and we can use the database instance as shown Figure~\ref{ex:decomposition:two}. For all other nodes (if any) in the join tree of $Q'$, as well as join tree of other connected components, we can create a relation per node with exactly one tuple such that $Q(D)$ is not empty and assign a uniform weight (say) $1$. Using the same argument as in \autoref{lem:lowerbound}, it is easy to see that any correct algorithm must examine the rank of $\Omega(n^2)$ tuples in order to find the smallest one since the tuple formed by the weights of the edges for any two output tuples will be incomparable.
    If there are at least two connected components, each with diameter two or more, then the join tree is of the form as shown in Figure~\ref{ex:decomposition:two} (with possibly more nodes in the join tree) with a root to leaf path of length at least two. Suppose $\mB_3$ and $\mB_2$ belong to one component and thus have a variable in common. Similarly, $\mB_1$ and $\mB_4$ also have a variable in common. Now, we can modify the database instance from Figure~\ref{fig:lowerbound} to have the schema for relations $R_2$ and $R_3$ as $R_2(y_1,z_1)$ and $R_3(y_2,z_2)$, and the domain of both $z_1$ and $z_2$ is $\{c\}$. Relations $R_1$ and $R_2$ correspond to $\mB_3$ and $\mB_2$, and  $R_3$ and $R_4$ correspond to $\mB_1$ and $\mB_4$. The tuples and weights from Figure~\ref{fig:lowerbound} remain the same. For all other nodes that may be in the join tree, we again create a relation with a single tuple, such that the output of the query is non-empty. Once again, we get $\Omega(n^2)$ tuples that are incomparable when looking at the weights of the edges that form the tuple. This completes the proof. \qedhere

    \eat{Consider the decomposition with the smallest depth and the least number of nodes. Consider the longest root to leaf path in such a decomposition. Our first observation is that any leaf node (for example, $\mB_3$ in Figure~\ref{ex:decomposition:two}) must have exactly one variable in common with its parent. Indeed, if there are no variables in common, then the leaf node can directly be made as a child of the parent of its parent. As an example, in Figure~\ref{ex:decomposition:two}, node for $\mB_3$ can directly be made as the child of the root if $\mB_2 \cap \mB_3 = \emptyset$. On the other hand, if both variables are identical, then the node can be removed from the tree. Observe that the same argument is applicable to any root to leaf path that has length two or more. Thus, the depth of the decomposition would be reduced.}

\begin{figure}[!t]
	\centering
	\scalebox{1}{\begin{tikzpicture}
		\tikzset{edge/.style = {->,> = latex'},
			vertex/.style={circle, thick, minimum size=7mm}}
		\def\x{0.25}
		
		\begin{scope}[fill opacity=1]
		\draw[fill=black!5] (0, 0) ellipse (1cm and 0.33cm) node {\small ${\color{black} x_1, x_2}$};
		\draw[fill=black!5] (-0.5, -1) ellipse (1cm and 0.33cm) node {\small ${\color{black} y_1, y_2}$};
		\draw[fill=black!5] (-1, -2) ellipse (1cm and 0.33cm) node {\small ${\color{black} z_1, z_2}$};
		\draw[fill=black!5] (2, -1) ellipse (1cm and 0.33cm) node {\small ${\color{black} w_1, w_2}$};
		\draw[edge] (0, -0.33) -- (-0.25,-.65);
		\draw[dotted] (0, -0.33) -- (-2,-0.65);
		\draw[dotted] (0, -0.33) -- (4,-0.65);
		\draw[edge] (-0.5, -1.33) -- (-0.75, -1.65);
		\draw[edge] (0, -0.33) -- (2, -0.65);
		
		\node[vertex]  at (-2, 0) {\small $\mB_{\texttt{root}} = \mB_1$};
		\node[vertex]  at (-2, -1) {\small $\mB_{2}$};				
		\node[vertex]  at (-2.5, -2) {\small $\mB_{3}$};	
		\node[vertex]  at (2, -1.66) {\small $\mB_{4}$};	
		\end{scope}	
		\end{tikzpicture}
	}
	\caption{Query decomposition example with depth more than one.}
	\label{ex:decomposition:two}
\end{figure}

\end{proof}
}

\noindent 
The results presented in this section demonstrate that small changes in the property of the ranking functions can lead to very different enumeration guarantees for the same query. For instance, for the cartesian product query $Q(x_1,y_1, x_2, y_2) = \ R(x_1,y_1), S(x_2, y_2)$, \autoref{lem:lowerbound} showed that no  linear preprocessing time and logarithmic delay algorithm can exist for coordinate-decomposable functions. However, \autoref{thm:edge:dichotomy} tells us that for edge-decomposable functions and the same query, there exists a linear preprocessing time and logarithmic delay algorithm. 

\hlrone{While we show dichotomies for full acyclic CQs over binary relations, a complete syntactic characterization for arbitrary full acyclic CQs remains an open problem. Our results for edge-decomposable and coordinate-decomposable ranking functions show that depending on the properties of the query hypergraph, a query may or may not admit efficient algorithms. However, no other ranking functions with reasonable restrictions are known that are \emph{intrinsically} hard. The problem of finding such  natural families of ranking function that are hard intrinsically and do not admit efficient enumeration algorithms any acyclic CQs is also interesting.}


\subsection{Beyond Logarithmic Delay} 
Next, we examine whether  the logarithmic factor that we obtain in the delay of \autoref{thm:main} can be removed for ranked enumeration. In other words, is it possible to achieve constant delay enumeration while keeping the preprocessing time small, even for simple ranking functions? To reason about this, we need to describe the {\em $X+Y$ sorting} problem.

Given two lists of $n$ numbers, $X = \langle x_1, x_2, \dots, x_{n} \rangle$ and $Y = \langle y_1, y_2, \dots, y_{n} \rangle$, we want to enumerate all $n^2$ pairs $(x_i, y_j)$ in ascending order of their sum $x_i + y_j$. This classic problem has a trivial $O(n^2 \log n)$ algorithm that materializes all $n^2$ pairs and sorts them. However, it remains an open problem whether the pairs can be enumerated faster in the RAM model. Fredman~\cite{fredman1976good}  showed that $O(n^2)$ comparisons suffice in the nonuniform linear decision tree model, but it remains open whether this can be converted into an $O(n^2)$-time algorithm in the real RAM model. Steiger and Streinu~\cite{steiger1995pseudo} gave a simple algorithm that takes $O(n^2 \log n)$ time while using only $O(n^2)$ comparisons. 

\begin{conj} [\cite{bremner2006necklaces, open}]
 {\em $X+Y$ sorting} does not admit an $O(n^2)$ time algorithm. \label{conjecture:sort:x+y}
\end{conj}

In our setting, $X+Y$ sorting can be expressed as enumerating the output of the cartesian product $Q(x, y) = R(x), S(y)$, where relations $R$ and $S$ correspond to the sets $X$ and $Y$ respectively. The ranking function is $\rf(x,y) = x + y$. Conjecture~\ref{conjecture:sort:x+y} implies that it is not possible to achieve constant delay for the cartesian product query and the sum ranking function; otherwise, a full enumeration would produce a sorted order in time $O(n^2)$.

	\section{Related Work}
\label{sec:related}

Top-k ranked enumeration of join queries has been studied extensively by the database community for both certain~\cite{li2005ranksql, qi2007sum, ilyas2004rank, li2005ranksql2} and uncertain databases~\cite{re2007efficient, zou2010finding}. Most of these works exploit the monotonicity property of scoring functions, building offline indexes and integrate the function into the cost model of the query optimizer in order to bound the number of operations required per answer tuple. We refer the reader to~\cite{ilyas2008survey} for a comprehensive survey of top-k processing techniques discovered prior to 2008. More recent work~\cite{chang2015optimal, gupta2014top} has focused on enumerating \emph{twig-pattern} queries over graphs. Our work departs from this line of work in two aspects: {\itshape(i)} use of novel techniques that use query decompositions and clever tricks to achieve strictly better space requirement and formal delay guarantees; {\itshape(ii)} our algorithms are applicable to arbitrary hypergraphs as compared to simple graph patterns over binary relations. Most closely related to our setting is~\cite{kimelfeld2006incrementally} and a line of work initiated by~\cite{yang2018any}.~\cite{kimelfeld2006incrementally} uses an adaptation of Lawler-Murty's procedure to incrementally computing ordered answers of full acyclic CQs. However, that work was mainly focused on studying the combined complexity of the problem. Further, since the goal was to obtain polynomial delay guarantees, the authors did not attempt to obtain the best possible delay guarantees. This line of work was further extended to parallel setting~\cite{golenberg2011optimizing} and also when the data is incomplete~\cite{kimelfeld2007combining}. 

\highlight{The other line of work was initiated by Yang et al.~\cite{yang2018any} who presented a novel anytime algorithm, called KARPET, for enumerating {\em homomorphic tree patterns} with worst case delay and space guarantees where the ranking function is sum of weights of input tuples that contribute to an output tuple. KARPET is an any-time algorithm that generates candidate output tuples with different scores and sorts them incremental via a priority queue. However, the candidate generation phase is expensive (which translates to linear delay guarantees) and can be improved substantially, as we show in this article.~\cite{yang2018anysig} made the further connection that KARPET can be extended to arbitrary full CQs (including cycles) by considering different tree decompositions. This connection was concretely established in concurrent work~\cite{tziavelis13optimal} that built upon~\cite{yang2018any, yang2018anysig} to obtain logarithmic delay guarantees using a dynamic programming approach combined with Lawler's procedure~\cite{lawler1972procedure}. \hlrtwo{Both our work and prior work~\cite{tziavelis13optimal} are generalizations of known algorithms~\cite{jimenez1999computing, eppstein1998finding} from paths to CQs. In comparison with~\cite{tziavelis13optimal}, we (1) present a framework that considers defines general properties of ranking functions and how to combine it with tree decompositions via the notion of compatibility, (2) we consider ranking functions beyond the sum of tuple weights as considered in~\cite{tziavelis13optimal}, and (3) present conditional and unconditional lower bounds. On the other hand,~\cite{tziavelis13optimal} considers CQs with projections (i.e., non-full CQs), conducts a thorough experimental evaluation on real-world datasets, and considers other measures of success such as \emph{time-to-k} $TT(k)$ which is defined as time required until the $k^{th}$ answer is returned. Note that a low delay is sufficient but not necessary to achieve low $TT(k)$.} More recently, the authors were also able to extend their results to theta-joins as well~\cite{tziavelis2021optimal}. For a more detailed overview of the prior work on the topic of ranked enumeration, we refer the reader to~\cite{tziavelis2020optimal, tziavelis13optimal}.}

\smallskip
\noindent \introparagraph{Rank aggregation algorithms} 	Top-k processing over ranked lists of objects has a rich history. The problem was first studied by Fagin et al.~\cite{fagin2002combining, fagin2003optimal} \hlrtwo{where the database consists of a \emph{single} relation $R(x_1, \dots, x_m)$ containing $N$ rows (referred to as objects) and $m$ attributes (referred to as ranked streams). The ranking function is defined over the the $m$ attributes and the goal is to find the top-$\mk$ results for coordinate monotone functions.} The authors proposed Fagin's algorithm (\textsc{FA}) and Threshold algorithm (\textsc{TA}), both of which were shown to be instance optimal for database access cost under  sorted list access and random access model. This model would be applicable to our setting only if $Q(D)$ is already computed and materialized \hlrtwo{(so as to obtain a single relation, which would be of size $|Q(D)|$)}.  More importantly, \textsc{TA} can only give $O(N)$ delay guarantee using $O(N)$ space.~\cite{natsev2001supporting} extended the problem setting to the case where we want to enumerate top-$\mk$ answers for $t$-path query. The first proposed algorithm $J^*$ uses an iterative deepening mechanism that pushes the most promising candidates into a priority queue. Unfortunately, even though the algorithm is instance optimal with respect to number of sorted access over each list, the delay guarantee is $\Omega(\vert Q(D) \vert)$ with space requirement $S = \Omega(\vert Q(D) \vert)$. A second proposed algorithm $J^*_{PA}$ allows random access over each sorted list. $J^*_{PA}$ uses a dynamic threshold to decide when to use random access over other lists to find joining tuples versus sorted access but does not improve formal guarantees. 

\smallskip
\noindent \introparagraph{Query enumeration} The notion of constant delay query enumeration was introduced by Bagan, Durand and Grandjean in~\cite{bagan2007acyclic}. In this setting, preprocessing time is supposed to be much smaller than the time needed to evaluate the query (usually, linear in the size of the database), and the delay between two output tuples may depend on the query, but not on the database. This notion captures the {\em intrinsic hardness} of query structure. For an introduction to this topic and an overview of the state-of-the-art we refer the reader to the survey~\cite{segoufin2013enumerating, segoufin2015constant}. Most of the results in existing works focus only on lexicographic enumeration of query results where the ordering of variables cannot be arbitrarily chosen. Transferring the static setting enumeration results to under updates has also been a subject of recent interest~\cite{berkholz2018answering, berkholz2017answering}.

\smallskip
\noindent \introparagraph{Factorized databases} Following the landmark result of~\cite{DBLP:journals/tods/OlteanuZ15} which introduced the notion of using the logical structure of the query for efficient join evaluation, a long line of research has benefited from its application to learning problems and broader classes of queries~\cite{bakibayev2012fdb, bakibayev2013aggregation, olteanu2016factorized, deep2018compressed, kara2020trade, deep2020fast, deep2021enumeration}. The core idea of factorized databases is to convert an arbitrary query into an acyclic query by finding a query decomposition of small width. This width parameter controls the space and pre-processing time required in order to build indexes allowing for constant delay enumeration. We build on top of factorized representations and integrate ranking functions in the framework to enable enumeration beyond lexicographic orders.

	\section{Conclusion}
\label{sec:conclusion}

In this paper, we study the problem of CQ result enumeration in ranked order. We combine the notion of query decompositions with certain desirable properties of ranking functions to enable logarithmic delay enumeration with small preprocessing time.  The most natural open problem is to prove space lower bounds to see if our algorithms are optimal at least for certain classes of CQs. An intriguing question is to explore the full continuum of time-space tradeoffs. For instance, for any compatible ranking function with the $4$-path query and $T_P = O(N)$, we can achieve $\delta = O(N^{3/2})$ with space $S_e = O(N)$ and $\delta = O(\log N)$ with space $S_e = O(N^2)$. The precise tradeoff between these two points and its generalization to arbitrary CQs is unknown. There also remain several open question regarding how the structure of ranking functions influences the efficiency of the algorithms. In particular, it would be interesting to find fine-grained classes of ranking functions which are more expressive than totally decomposable, but less expressive than coordinate decomposable. For instance, the ranking function $f(x,y) = \vert x - y \vert$ is not coordinate decomposable, but it is {\em piecewise} coordinate decomposable on either side of the global minimum critical point for each $x$ valuation.  \highlight{Finally, recent work has made considerable progress in query evaluation under updates. In this setting, the goal is to minimize the update time of the data structure as well as minimize the delay. A simple application of our algorithm is useful here. For any full acyclic query, one can maintain the relations under updates in constant time by updating the hash maps and then apply the preprocessing and enumeration phase of our algorithm. This algorithm gives a linear delay guarantee since the preprocessing phase takes linear time. One could also apply the preprocessing phase of our algorithm after each update to reset all priority queues which makes the update time linear but the enumeration delay can now be $O(\log|D|)$. Both of these guarantees can be improved upon for the class of hierarchical queries~\cite{berkholz2017answering, kara2020trade}. We leave the precise construction, algorithms, and empirical evaluation as a topic for future research.}

    \newpage
	\bibliographystyle{alphaurl}
	\bibliography{reference}
	
\end{document}